 \documentclass[final,5p,times,twocolumn]{elsarticle}

\usepackage{lineno,hyperref}
\usepackage{graphicx}
\usepackage{algorithm} 
\usepackage{algpseudocode}
\usepackage{amssymb,amsfonts}
\usepackage{amsmath}
\usepackage{xcolor}
\usepackage{longtable,pdflscape}
\usepackage{booktabs}
\usepackage{supertabular}

\usepackage{graphicx}
\usepackage{subfigure}
\usepackage{dcolumn}
\usepackage{bm}
\usepackage[T1]{fontenc}
\usepackage{mathptmx}
\usepackage{multirow}
\usepackage{array}
\usepackage{float}
\usepackage{lscape}
\modulolinenumbers[5]

\journal{Computers and fluids}

\begin{document}

\begin{frontmatter}



\title{Specific-heat ratio effects on the interaction between shock wave and heavy-cylindrical bubble: based on discrete Boltzmann method}

 \author[label1,label2]{Dejia Zhang}
 \address[label1]{State Key Laboratory for GeoMechanics and Deep Underground Engineering, China University of Mining and Technology, Beijing 100083, P.R.China}
 \address[label2]{Laboratory of Computational Physics, Institute of Applied Physics and Computational Mathematics, P. O. Box 8009-26, Beijing 100088, P.R.China}

 \author[label2,label4,label5]{Aiguo Xu\corref{mycorrespondingauthor}}
\address[label4]{HEDPS, Center for Applied Physics and Technology, and College of Engineering, Peking University, Beijing 100871,P.R.China}
\address[label5]{State Key Laboratory of Explosion Science and Technology, Beijing Institute of Technology, Beijing 100081, P.R.China}
\cortext[mycorrespondingauthor]{Corresponding author at: Laboratory of Computational Physics, Institute of Applied Physics and Computational Mathematics, P. O. Box 8009-26, Beijing 100088, P.R.China.}
\ead{Xu\_Aiguo@iapcm.ac.cn}

\author[label6,label2]{Jiahui Song }
\address[label6]{School of Aerospace Engineering, Beijing Institute of Technology, Beijing, 100081, P.R.China}

\author[label7]{Yanbiao Gan }
\address[label7]{Hebei Key Laboratory of Trans-Media Aerial Underwater Vehicle, School of Liberal Arts and Sciences, North China Institute of Aerospace Engineering, Langfang 065000, P.R.China}

\author[label8]{Yudong Zhang}
\address[label8]{School of Mechanics and Safety Engineering, Zhengzhou University, Zhengzhou 450001, P.R.China}

\author[label1]{Yingjun Li \corref{mycorrespondingauthor}}
\cortext[mycorrespondingauthor2]{State Key Laboratory for GeoMechanics and Deep Underground Engineering, China University of Mining and Technology, Beijing 100083, P.R.China}
\ead{ lyj@aphy.iphy.ac.cn}

\author{}

\address{}

\begin{abstract}
\textcolor{red}{Specific-heat ratio effects on the interaction between a planar shock wave and a two-dimensional heavy-cylindrical bubble are studied by the discrete Boltzmann method.
Snapshots of schlieren images and evolutions of characteristic scales, being consistent with experiments, are obtained.
The specific-heat ratio effects on some relevant dynamic behaviors  such as the bubble shape, deformation process, average motion, vortex motion, mixing degree of the fluid system are carefully studied, as well as the related Thermodynamic Non-Equilibriums (TNE) behaviors including the TNE strength, entropy production rate of the system.
Specifically, it is found that the influence of specific-heat ratio on the entropy production contributed by non-organized energy flux (NOEF) is more significant than that caused by non-organized momentum flux (NOMF).
Effects of specific-heat ratio on entropy production caused by NOMF and NOEF are contrary.
The effects of specific-heat ratio on various TNE quantities show interesting differences.
These differences consistently show the complexity of TNE flows which is still far from clear understanding.}
\end{abstract}


\end{frontmatter}


\section{\label{sec:level1} Introduction}
The applications of shock-accelerated inhomogeneous flows (SAIFs) are of significant value in biomedicine, energy utilization, and astrophysics fields, including but not limited to scenarios such as the impact of shock waves on kidney stones, the interaction between shock waves with foams, the impacting of detonation wave with burning flames in supersonic combustion systems, the formation of supernova remnants, etc \cite{Ranjan2011ARFM,Zou2017PRE,Liu2022POP,2017Lin-DDBM-RT,Chen2016FOP,Cai2021MRE,Saurel2003JFM,Singh2021POF}.
Shock-bubble interaction (SBI) is one of the most fundamental problems in the research of SAIFs.
Its applications and academic research are interdisciplinary.
Generally, there are two kinds of problems encountered in SBI research:
(i) The geometry of shock waves, the shape of material interfaces, and the structure of container are complex in the actual scene.
They will result in various wave patterns and significantly affect the flow morphology and bubble's evolution.
(ii) There usually exist multi-physics coupling problem in the engineering application of SBI.
Such as the supersonic combustion machines.
When the shock waves passing through the reactants, it may lead to phase transition and chemical reactions, making the flow morphology more complex and inducing small structure (or fast-changing pattern) \cite{Haehn2012CNF,Diegelmann2017CNF,Fan2022CNF}.
In an underwater explosion experiment, the interaction between shock waves and bubbles may refer to the cavitation and annihilation effects.
The other scene is the inertial confinement fusion (ICF), in which the laser ablation, electron heat conduction, self-generated electromagnetic field, radiation, and many other factors may complicate the investigation of hydrodynamic instabilities \cite{Rinderknecht2018kinetic}.

Commonly, research on SBI mainly includes three methods: theoretical derivation, experiment, and numerical simulation.
As a fundamental research method, theoretical research can provide a clear understanding of physical processes.
In 1960, Rudinger \emph{et al.} developed a theory that permits computing the response of bubbles to accelerations \cite{Rudinger1960JFM}.
In order to describe the formation and evolution processes of vortex structure quantitatively, many scholars have developed circulation models \cite{Picone1988JFM,Yang1994JFM,Samtaney1994JFM,Niederhaus2008JFM}.
However, theoretical works provide limited information.
Meanwhile, in the late stage of SBI evolution, the bubble deformation and flow morphology dominated by the developed Richtmyer-Meshkov instability (RMI) and Kelvin-Helmholtz instability (KHI) are difficult to be predicted accurately.

As the research method closest to engineering application, the experimental results are often regarded as standard results to verify the rationality and accuracy of theoretical and numerical works.
To study the SBI process accurately, the scholars have made a series of improvements to experimental equipment or technique, including the generation techniques of different types of shock waves, interface formation methods, schlieren facilities, and image recognition techniques \cite{Haas1987JFM,Jacobs1992JFM,
Layes2003PRL,Layes2005POF,Tomkins2008JFM,Ranjan2005PRL,Ranjan2007PRL,
Zhai2011POF,Ding2017PRL,Zhai2018Review,Ding2017JFM,Liang2018SCPMA,
Luo2019JFM2,Ding2018POF}.
Among these, two of important and valuable works are performed by Ding \emph{et al.}.
Based on the soap film technique, they formed kinds of initial interfaces with different curvatures through the wire-restriction method and captured the wave patterns and interface evolution with high-speed schlieren photography \cite{Ding2017JFM,Ding2018POF}.
Other works, such as evolutions of a spherical gas interface under reshock conditions \cite{Si2012POF}, developments of a membrane-less $\rm{SF_{6}}$ gas cylinder under reshock conditions \cite{Zhai2014JV}, and interactions of a cylindrical converging shock wave with an initially perturbed gaseous interface \cite{Si2014LPB}, are also performed by many other scholars.

However, we know that the experimental studies mainly depend on the experimental platform.
When studying some complex and demanding condition problems, it takes much work to build the experimental platform.
In this situation, numerical simulation research becomes an option.
Generally, there are three kinds of physical modeling methods (or models) for SBI numerical research, i.e., the macroscopic, mesoscopic, and microscopic modeling methods.
Most of the existing numerical researches on SBI are related to the macroscopic modeling methods (such as the Euler and Navier-Stokes (NS) models) based on the continuous hypothesis (or equilibrium and near-equilibrium hypothesis)\cite{Picone1988JFM,Yang1994JFM,Quirk1996JFM,Samtaney1994JFM,Giordano2006POF,
Niederhaus2008JFM,Zou2019JFM,Zou2016POF,Sha2015ActaPS,li2019POF,
Chen2021POF,Hejazialhosseini2013POF,Zhu2017POF,Liu2022JFM,Wang2015POF,
Wang2018POF,Yu2021PRF,Yu2020POF,Ding2018POF,Zhai2011POF}.
 For example, Zou \emph{et al.} presented the computational results on the evolution of the shock-accelerated heavy bubbles through the multi-fluid Eulerian equation \cite{Zou2015SCPMA}.
There also exist a few SBI works based on the mesoscopic modeling method, such as
the Direct Simulation Monte Carlo method \cite{Zhang2019CNF}.
The microscopic modeling methods such as the Molecular dynamics (MD) simulation, is capable of capturing much more flow behaviors but restricted to smaller spatiotemporal scales because of its huge computing costs.

In the numerical research on SBI, three points need to be concerned.
(i) Investigation of kinetic modeling that describes the non-continuity/non-equilibrium flows.
Most of the current researches are based on macroscopic models.
However, there exist abundant small structure (and fast-changing patterns) behaviors and effects such as the shock wave, boundary layer, material defects, etc.
For cases with small structures, the mean free path of molecules cannot be ignored compared to the characteristic length, i.e., the non-continuity (discreteness) of the system is pronounced,
which challenge the rationality and physical function of the macroscopic models based on the continuity hypothesis.
For cases with fast-changing patterns, the system dose not have enough time to relax to the thermodynamic equilibrium state, i.e., the system may significantly deviate from the thermodynamic equilibrium state.
Therefore, the rationality and physical function of the macroscopic models based on the hypothesis of thermodynamic equilibrium (or near thermodynamic equilibrium) will be challenged.
(ii) Improvement of method that describes the evolution characteristics of bubbles and flows morphology.
Most of the studies describe bubble characteristics and flows morphology from a macroscopic view.
The mesoscopic characteristics such as the kinetic effects which help understand the kinetic process, are rarely to be studied.
(iii) Further studies of effects of specific-heat ratio on SBI process.
The specific-heat ratio is an essential index for studying the compressibility of the gas.
Research from Igra \emph{et al.} has shown that the differences in the specific-heat ratio of bubbles would cause various wave patterns and pressure distribution inside the bubbles during the interaction process \cite{Igra2018POF}.
Besides, many works on hydrodynamic instability have also demonstrated the importance of investigating the specific-heat ratio effect \cite{Livescu2004POF,Xue2010POP,Fries2021JFM,Chen2021FOP}.
Among these, Chen \emph{et al.} investigated the specific-heat ratio effects on temperature gradient and the TNE characteristics of compressible Rayleigh-Taylor (RT) system \cite{Chen2021FOP}.

\textcolor{blue}{For the above three points, in this work we  \textcolor{red}{apply} the recently proposed discrete Boltzmann method (DBM)
\cite{Xu2022BSP,Zhang2022POF,Gan2022JFM,2022Zhang-AIPAdv-Slip,Xu2012-FoP-review,2015Xu-PRE,2018Gan-pre,Zhang2017Discrete,Xu2018-Chap2,2021XuACTAA,2021XuCJCP,2021XuACTA}.}
\textcolor{red}{The Lattice Boltzmann Method (LBM) research has two complementary branches\cite{2001-Succi-Boltzmann,Qian1992EL,Guo2015PR,Sun2016IJHMT,Sun2016IJHMT2, Xu2022BSP,Zhang2022POF,Gan2022JFM}.
One aims to work as a kind of new scheme for numerical solving various partial differential equation(s). The other aims to work as a kind of new method for constructing  kinetic model to bridge the macro and micro descriptions. The two branches have different goals and consequently have different rules. The current DBM is developed from the second branch of LBM and focusing more on the Thermodynamic Non-Equilibrium (TNE) behaviors that the macro modeling generally ignore.  It breaks through the continuity and near-equilibrium assumptions of traditional fluid modeling, discards the lattice gas image of standard LBM, and adds various  methods based on phase space for checking, exhibiting, describing and analyzing the non-equilibrium state and resulting effects. More information extraction technologies and analysis methods for complex field are introduced with time.}

\begin{figure*}[htbp]
\center\includegraphics*
[width=0.8\textwidth]{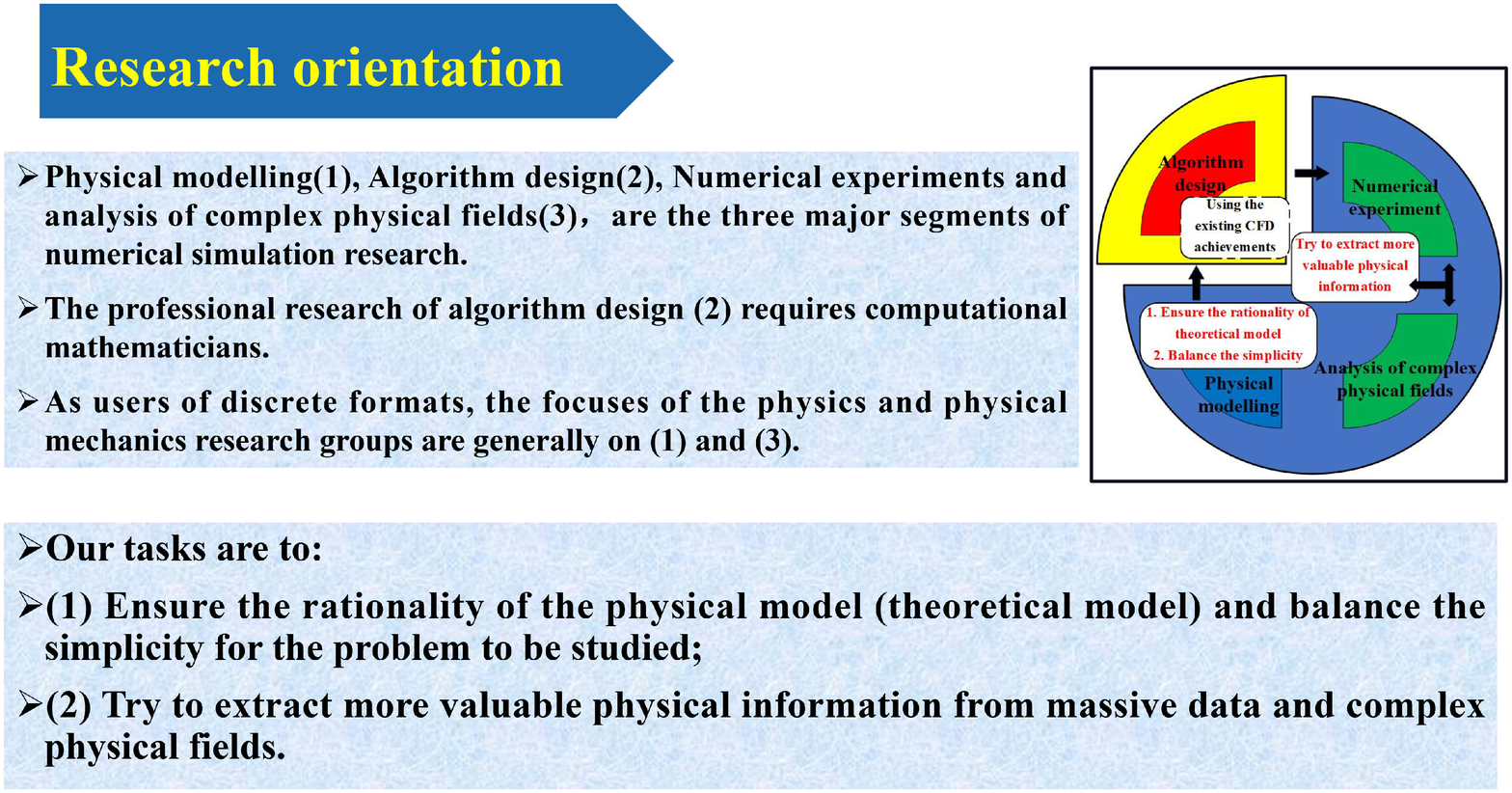}
\caption{  Research orientation and tasks of DBM.} \label{fig}
\end{figure*}

\textcolor{red}{The numerical simulation includes three parts, as shown in Fig. \ref{fig}.
(1) Physical modelling, (2) Algorithm design, (3) Numerical experiments and analysis of complex physical fields.
The research of equation algorithm corresponds to the part (2) of the above three parts.
The DBM aims at parts (1) and (3) of the three mentioned above.
It belongs to a physical model construction method rather than a numerical solution for the equations.
The tasks of DBM are to:
(i) Ensure the rationality of the physical model (theoretical model) and balance the simplicity for the problem to be studied;
(ii) Try to extract more valuable physical information from massive data and complex physical fields.}

Based on the coarse-grained modeling method
of non-equilibrium statistical physics, the DBM aims to solve the following dilemma:
(i) The traditional hydrodynamic modelings are based on the continuous hypothesis (or near-equilibrium hypothesis).
They only concern the evolution of three conserved kinetic moments of the distribution function, i.e. the density, momentum and energy,
so their physical functions are insufficient.
(ii) The situation that the MD can be used is restricted to too small spatial-temporal scales.
The physical requirement for the modeling is that
except for the Hydrodynamic Non-Equilibriums (HNE), the most related TNE are also needed to be captured.
Theoretically, the Boltzmann equation is suitable for all-regime flows, including the continuum regime, slip regime, transition regime, and free molecule flow regime.
Based on the Chapman-Enskog (CE) multiscale analysis \cite{1990Chapman}, through retaining various orders of Kn number (or considering different orders of TNE effects), the Boltzmann equation can be reduced to the various orders of hydrodynamic equations.
They can be used to describe the hydrodynamic behaviors, i.e., the conservations of mass, momentum and energy
, in corresponding flow regimes. Because what the traditional hydrodynamic equations describe are only the conservation laws of mass, momentum and energy.
Consequently, it should be pointed out that, the information lost in the traditional hydrodynamic equations increases sharply with increasing the Kn number.
With increasing the Kn number, to ensure the describing capability not to decrease significantly, the more appropriate hydrodynamic equations should be the Extended Hydrodynamic Equations (EHEs) which include not only the evolution equations of conserved kinetic moments  but also the most relevant non-conserved kinetic moments of distribution function.
For convenience of description we refer the modeling method that derives EHEs from the fundamental kinetic equation to Kinetic Macroscopic Modeling (KMM) method.
It is clear that, the complex process of CE expansion is necessary and the simulation is still based on the macroscopic equations in KMM.
As a comparison,
the DBM is a kind of Kinetic Direct Modeling (KDM) method.
In DBM modeling, the CE analysis is only used to quickly determine which kinetic moments should keep values unchanged, the final EHEs are not needed, and the simulation is not based on the complicated EHEs.
As the TNE degree of the flow to be described rises gradually, the complexity of the derivation process and difficulty of numerical simulation in the KMM method increase sharply.
However, in the DBM method, to describe flows in a one-order more deeper depth of TNE, only two more related kinetic moments need to be added. Since without needing to derive and solve the EHEs,
as the TNE degree deepens, the complexity of the DBM approach increases much slower than that of KMM method.

The core step in DBM modeling is to provide a feasible scheme for detecting, describing, presenting, and analyzing TNE effects and behaviors beyond traditional macroscopic modeling.
Based on the non-equilibrium statistical physics, we can use the non-conservative moments of ($f-f^{eq}$) to describe how and how much the system deviates from the thermodynamic equilibrium state and to check corresponding effects due to deviating from the thermodynamic equilibrium.
The non-conservative moments of ($f-f^{eq}$) open a phase space, and this space and its subspaces provide an intuitive geometric correspondence for describing complex TNE system properties.
The development of schemes for checking TNE state, extracting TNE information and describing corresponding TNE effects in DBM are seen in Table \ref{Table1}.
Actually, this set of TNE describing methods has been applied in many kinds of complex fluid systems such as hydrodynamic instability system \cite{Lai2016PRE,2017Lin-DDBM-RT,Chen2018POF,Ye2020Entropy,Lin2021PRE,Zhang2021POF,Chen2022FOP,Chen2022PRE},
combustion and detonation systems \cite{2015Xu-PRE,Lin2016CNF,Lin2018CAF,Lin2018CNF,Ji2022JCP,Shan2022JMES,2021XuACTAA}, multiphase flow system \cite{Gan2011PRE,Gan2015Soft,Zhang2019Matter,Zhang2020FOP,Gan2022JFM,
Gan2019FOP}, plasma system \cite{Liu2022JMES}, etc.
Besides the scheme for detecting, describing, presenting, and analyzing TNE effects and behaviors, the DBM incorporates other methods for analyzing the complex physical field.
One of them is the tracer particle method.
The introduction of the tracer particle method makes the gradually blurred interface appear clearly \cite{Zhang2021POF,Li2022CTP}.
\begin{center}
\begin{table*}
\begin{tabular}{p{2cm}<{\centering} p{15cm}<{\centering}}
\hline
\hline
Year    &Scheme for investigating TNE effects and behaviors \\
\hline
Before 2012    &Two classes of LBMs did not show a significant difference in physical function. \\
2012 &    Use the non-conservative moments of ($f-f^{eq}$) to check and describe TNE  \cite{Xu2012-FoP-review}.
This is the starting point of current DBM approach.
 \\
2015 & Open TNE phase space based on non-conservative moments of ($f-f^{eq}$) and define a TNE strength using the distance from a state point to the origin.  This is the starting point of the phase space description method \cite{2015Xu-PRE}.
 \\
2018 & Extend the distance concepts in phase space to describe the TNE difference/similarity of TNE states and kinetic processes \cite{Xu2018-RGD31}. \\
2021 & Further extend the phase space description methodology to any set of system characteristics \cite{2021XuCJCP}.  \\
\hline
\hline
\end{tabular}
\caption{The development of schemes for checking TNE state, extracting TNE information and describing corresponding TNE effects in DBM.}
\label{Table1}
\end{table*}
\end{center}

The rest of the paper is structured as follows.
Section \ref{Model construction} shows the modeling method.
Then, the numerical simulations and results are presented in Section \ref{Numerical simulations}, which includes two subsections.
Section \ref{Conclusions} concludes the paper.
Other complementary information is given in the Appendix.

\section{Model construction }\label{Model construction}
Based on the Bhatnagar-Gross-Krook (BGK) single-relaxation model, a two-fluid DBM with a flexible specific-heat ratio is presented in this part.
From the origin Boltzmann equation to a DBM, four fundamental steps are needed:
(i) Simplification and modification of the Boltzmann equation according to the research requirement.
(ii) Discretization of the particle velocity space under the condition that the reserved kinetic moments keep their values unchanged.
(iii) Checking the TNE state and extracting TNE information.
(iv) The selection/design of the boundary conditions.

\subsection{ Simplification and modification of the Boltzmann equation }

As we know, the collision term in the original Boltzmann contains high dimensional distribution functions.
Therefore, the direct solution to it needs too much computing consumption.
The most common method to simplify the collision operator is to introduce a local equilibrium distribution function ($f^{eq}$) and write the complex collision operator in a linearized form, i.e., the original BGK collision operator $-\frac{1}{\tau}(f-f^{eq})$, where $\tau$ is the relaxation time \cite{Bhatnagar1954BGK}.
The original BGK operator describes the situation where the system is always in the quasi-equilibrium state.
Namely, it characterizes only the situation where the Kn number of the system is small enough and $f \approx f^{eq}$.
The currently used BGK operator for non-equilibrium flows in the field is a modified version incorporating the mean-field theory description \cite{Xu2022BSP,Zhang2022POF,Gan2022JFM}.
Based on the above considerations, the simplified Boltzmann equation describing the SBI process is
\begin{equation}
\frac{\partial f}{\partial t}+\mathbf{v}\cdot\frac{\partial f}{\partial \mathbf{r}}=-\frac{1}{\tau}(f-f^{eq})
,
\end{equation}
where the two-dimensional equilibrium distribution function is
\begin{equation}
\begin{aligned}
f^{eq}=\frac{\rho}{2\pi RT}(\frac{1}{2\pi IRT})^\frac{1}{2} \exp[-\frac{(\mathbf{v}-\mathbf{u})^2}{2RT}-\frac{\eta^{2}}{2IRT}]
\label{Eq.fES}
,
\end{aligned}
\end{equation}
where $\rho$, $T$, $\mathbf{v}$, $\mathbf{u}$, $I$, $R$, and $\eta$ are the mass density, temperature, particle velocity vector, flow velocity vector, the number of the extra degrees of freedom including molecular rotation and vibration inside the molecules, gas constant, and a free parameter that describes the energy of the extra degrees of freedom, respectively.
The specific-heat ratio is flexible by adjusting parameter $I$, i.e., $\gamma=(D+I+2)/(D+I)$, where $D=2$ represents the two-dimensional space.

\subsection{  Discretization of the particle velocity space and determination of $f^{\sigma,eq}_{i}$ }

The continuous Boltzmann equation should be discretized for simulating.
Specifically, the continuous velocity space can be replaced by a limited number of particle velocities.
So that the values of continuous kinetic moments can be obtained from the summation form of kinetic moments.
In this process, it requires the reserved kinetic moments, which are used to characterize the system behaviors, keep their values unchanged after discretizing the velocity space.
Namely, $\int f \Psi '(\mathbf{v}) d\mathbf{v}=\sum_{i} f_i \Psi '(\mathbf{v}_i)$,
where $\Psi ' = [1,\mathbf{v},\mathbf{vv},\mathbf{v\cdot v},\mathbf{vvv},\mathbf{vv\cdot v},\cdots]^{T}$ represent the reserved kinetic moments.
According to the CE analysis, $f$ can be expressed by $f^{eq}$.
Therefore, in the process of discretization, the reserved kinetic moments of $f^{eq}$ should keep their values unchanged, i.e., $\int f^{eq} \Psi ''(\mathbf{v}) d\mathbf{v}=\sum_{i} f_i^{eq} \Psi ''(\mathbf{v}_i)$.

 \textcolor{red}{The discrete Boltzmann is}
\begin{equation}
\frac{\partial f_{i}}{\partial t}+v_{i\alpha}\cdot\frac{\partial f_{i}}{\partial r_{\alpha}}=-\frac{1}{\tau}(f_{i}-f^{eq}_{i})
\label{Eq.Discrete-Boltzmann}
,
\end{equation}
where $i$ represents the kind of discrete velocities and $\alpha$ ($\alpha=x\;$ or $\;y$) is the direction in cartesian coordinate.

To simulate the interaction between two different fluids, a two-fluid DBM should be constructed.
Based on the single-relaxation model,  \textcolor{red}{the discrete two-fluid Boltzmann equation can be written as \cite{Zhang2020POF}:}
\begin{equation}
\frac{\partial f_{i}^{\sigma}}{\partial t}+v_{i\alpha}\cdot\frac{\partial f_{i}^{\sigma}}{\partial r_{\alpha}}=-\frac{1}{\tau^{\sigma}}(f_{i}^{\sigma}-f^{\sigma,eq}_{i})
\label{Eq.Discrete-Boltzmann1}
,
\end{equation}
where $\sigma$ represents the types of material particle and $f^{\sigma,eq}_{i}=f^{\sigma,eq}_{i}(\rho^{\sigma},\bf{u},\textit{T})$.
In two-fluid DBM, the macroscopic quantities of the mixture and each component are
\begin{equation}
\rho^{\sigma}=\sum_{i}f^{\sigma}_{i},
\end{equation}
\begin{equation}
\mathbf{u}^{\sigma}=\frac{\sum_{i}f^{\sigma}_{i}\mathbf{v}_{i}}{\rho^{\sigma}} ,
\end{equation}
\begin{equation}
\rho=\sum_{\sigma}\rho^{\sigma} ,
\end{equation}
\begin{equation}
\mathbf{u}=\frac{\sum_{\sigma}\rho^{\sigma}\mathbf{u}^{\sigma}}{\rho} ,
\end{equation}
where $\rho^{\sigma}$ and $\mathbf{u}^{\sigma}$ are the mass density and flow velocity of the component $\sigma$, respectively.
$\rho$ and $\mathbf{u}$ represent the mass density and flow velocity of the mixture, respectively.
\textcolor{red}{There exist two kinds of temperature (internal energy) definitions in two-fluid DBM because the definition of temperature (internal energy) depends on the flow velocity to be chosen as a reference.
The first definition is by choosing the velocity of the mixture to be a reference, i.e., $E^{\sigma*}_{I}=\frac{1}{2}\sum\limits_{i}f^{\sigma}_{i}((\mathbf{v}_{i}-\mathbf{u})^2+\eta^{\sigma 2}_{i})$.
So that the expressions of temperature of component $\sigma$ and mixture are
\begin{equation}
T^{\sigma*}=\frac{2E_{I}^{\sigma*}}{\rho^{\sigma}(D+I^{\sigma})} ,
\end{equation}
\begin{equation}
T=\frac{2E_{I}^{*}}{\sum_{\sigma}\rho^{\sigma} (D+I^{\sigma})}
,
\end{equation}
where $E_{I}^{*}=\sum_{\sigma}E_{I}^{\sigma*}$.
We can also choose the flow velocity of component as a reference, i.e., $E^{\sigma}_{I}=\frac{1}{2}\sum\limits_{i}f^{\sigma}_{i}((\mathbf{v}_{i}-\mathbf{u}^{\sigma})^2+\eta^{2}_{i})$, where $\mathbf{u}^{\sigma}$ is the flow velocity of component $\sigma$.
The corresponding definitions of temperature for component $\sigma$ and the mixture are
\begin{equation}
T^{\sigma}=\frac{2E_{I}^{\sigma}}{\rho^{\sigma}(D+I^{\sigma})} ,
\end{equation}
\begin{equation}
T=\frac{2(E_{I}+\Delta E_{I}^{*})}{\sum_{\sigma}\rho^{\sigma} (D+I^{\sigma})}
,
\end{equation}
where $\Delta E_{I}^{*}$ is
\begin{equation}
\Delta E_{I}^{*}=E_{I}^{*}-E_{I}=\frac{\rho^{A}\rho^{B}(u^{A}_{\alpha}-u^{B}_{\alpha})^{2}}{2(\rho^{A}+\rho^{B})} .
\end{equation}
It is clear to see that these two definitions of temperature for mixture are the same, but for temperature of component $\sigma$ are different.
We choose the first definition in this manuscript.}

To solve the Eq. (\ref{Eq.Discrete-Boltzmann1}), it is necessary to determine the values of $f^{\sigma,eq}_{i}$.
Its values depend on the reserved kinetic moments which characterize the main system behaviors.
In DBM modeling, the CE multiscale analysis is used to determine quickly the reserved kinetic moments.
Specifically, when constructing a DBM which only the first order term of Kn number is retained (i.e., only the first order TNE effects are retained), seven kinetic moments should be reserved, i.e., the $\mathbf{M}_0$, $\mathbf{M}_1$, $\mathbf{M}_{2,0}$, $\mathbf{M}_2$, $\mathbf{M}_{3,1}$, $\mathbf{M}_3$, $\mathbf{M}_{4,2}$.
Two more kinetic moments ( $\mathbf{M}_4$ and $\mathbf{M}_{5,3}$) are needed when the second order TNE is considered \cite{Zhang2022POF}.
However, it should be noted that the function of CE analysis in DBM modeling is only to determine the kinetic moments that need to be preserved.
Whether or not to derive the hydrodynamic equations does not affect the DBM simulation.
The kinetic moments used in our physical modeling are shown in the \ref{sec:AppendixesB}.
Their expressions can be obtained by integrating $\mathbf{v}$ and $\eta$ with continuous-form $f^{eq}$.
For better understanding, the \ref{sec:AppendixesC} gives the two-fluid hydrodynamic equations recovered from the Boltzmann equation.

\begin{figure}[htbp]
\center\includegraphics*
[width=0.4\textwidth]{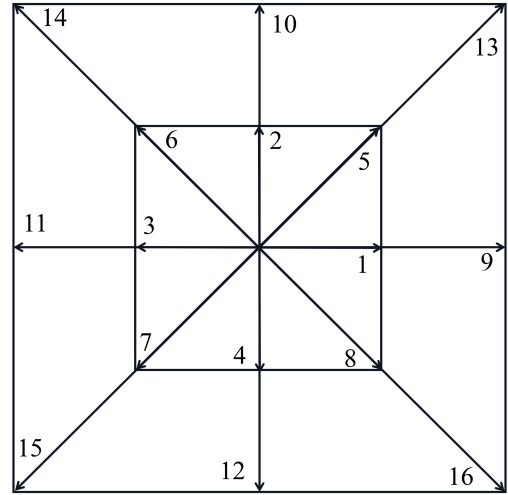}
\caption{Sketch of D2V16 model.
The numbers in the figure represent the index $i$ in Eq. (\ref{Eq.Discrete-Boltzmann}).} \label{fig1}
\end{figure}

The kinetic moments in \ref{sec:AppendixesB} can be written in matrix form, i.e.,
\begin{equation}
\mathbf{C}\cdot\mathbf{f}^{\sigma,eq}=\mathbf{\hat{f}}^{\sigma,eq}
,
\end{equation}
where $\mathbf{C}$ is the matrix of discrete velocity and $\mathbf{\hat{f}}^{eq}$ represents the kinetic moments.
A proper discrete velocity model is needed to confirm the values of $f^{\sigma, eq}_{i}$.
The $\mathbf{f}^{\sigma, eq}$ can be obtained by solving the inverse matrix, i.e., $\mathbf{f}^{\sigma, eq}=\mathbf{C}^{-1}\cdot\mathbf{\hat{f}}^{\sigma, eq}$, where $\mathbf{C}^{-1}$ is the inverse matrix of $\mathbf{C}$.
It is very convenient to obtain the inverse matrix of $\mathbf{C}$ through some mathematical softwares such as Mathematica, etc.
The D2V16 model is chosen in this paper, its sketches can be seen in Fig. \ref{fig1}.
The specific values of D2V16 are given in the following equations:
\[
(v_{ix},v_{iy})=
\left\{
\begin{array}{lll}
$\rm{cyc}$: c(\pm 1,0), &i& = 1-4 , \\
c(\pm1,\pm 1), &i& = 5-8 , \\
$\rm{cyc}$: 2c(\pm 1,0), &i& = 9-12 , \\
2c(\pm1,\pm 1), &i& = 13-16 ,
\end{array} \label{Eq:DDBM-DVM1}
\right.
\]
where ``cyc'' indicates cyclic permutation and $c$ is an adjustable parameter of the discrete velocity model.
The sketch of $\eta$ in D2V16 is $\eta_{i}=\eta_{0}$ for $i=1-4$, and $\eta_{i}=0$ for $i=5-16$.

\subsection{ Checking the TNE state and extracting TNE information  }

Many physical quantities can characterize the degree of TNE in a fluid system, such as relaxation time, Kn number, viscosity, heat conduction, the gradients of macroscopic quantity, etc.
They are all helpful to characterize the TNE strength and describe the TNE behaviors of a fluid system from their perspectives.
But it is not enough only relying on these quantities.
Besides the above physical quantities describing the TNE behaviors, in DBM modeling, we can also use the non-conservative moments of ($f-f^{eq}$) to characterize the TNE state and extract TNE information from the fluid system.

Fundamentally, four TNE quantities can be defined in a first-order DBM, i.e., $\bm{\Delta}^{\sigma*}_{2}$, $\bm{\Delta}^{\sigma*}_{3,1}$, $\bm{\Delta}^{\sigma*}_{3}$, and $\bm{\Delta}^{\sigma*}_{4,2}$.
Their definitions can be seen in Table \ref{Table2},
where $\mathbf{v}^{*}_{i}=\mathbf{v}_{i}-\mathbf{u}$ represents the central velocity and  $\mathbf{u}$ is the macro flow velocity of the mixture.
Physically, $\bm{\Delta^{\sigma*}_{2}}=\Delta^{\sigma*}_{2,\alpha\beta}\mathbf{e}_{\alpha}\mathbf{e}_{\beta}$ and $\bm{\Delta}^{\sigma*}_{3,1}=\Delta^{\sigma*}_{3,1}\mathbf{e}_{\alpha}$ represent the viscous stress tensor (or non-organized momentum flux, NOMF) and heat flux tensor (or non-organized energy flux, NOEF), respectively. The $\mathbf{e}_{\alpha}$ ($\mathbf{e}_{\beta}$) is the unit vector in the $\alpha$ ($\beta$) direction.
The later two higher-order TNE quantities contain more condensed information.
Specifically, $\bm{\Delta^{\sigma*}_{m,n}}$ ($\bm{\Delta^{\sigma*}_{m}}$) is the flux of $\bm{\Delta^{\sigma*}_{m-1,n-1}}$ ($\bm{\Delta^{\sigma*}_{m-1}}$).
For example,  $\bm{\Delta^{\sigma*}_{3}}$ is the flux of  $\bm{\Delta^{\sigma*}_{2}}$ and it indicates the flux information of $\bm{\Delta^{\sigma*}_{2}}$.
To describe the TNE strength of the whole fluid system, some TNE quantities contained more condensed information are also defined, i.e., $D^{\sigma*}_{2},D^{\sigma*}_{3,1},D^{\sigma*}_{3},D^{\sigma*}_{4,2},D^{*}_{m}, D^{*}_{m,n}$.
Other TNE quantities can be defined based on specific requirements.
All the independent components of TNE characteristic quantities open a high-dimensional phase space, and this space and its subspaces provide an intuitive image for characterizing the TNE state and understanding TNE behaviors \cite{Li2022CTP,Chen2022PRE,Zhang2022POF}.
It should be emphasized that:
(i) The TNE strength/intensity/degree is the most basic parameter of non-equilibrium flow description;
And any definition of non-equilibrium strength/intensity/degree  depends on the research perspective.
(ii) The physical meaning of $D^{*}_{m,n}$ is the TNE strength of this perspective.
(iii) From a certain perspective, the TNE strength is increasing; While from a different perspective, the TNE strength, on the other hand, may be decreasing. It is normal, one of the concrete manifestations of the complexity of non-equilibrium flow behavior.
Strictly speaking, those TNE intensity and effect descriptions that do not account for the research perspective are not correct.
Do not explain the research perspective, the corresponding is not dependent on the research perspective.
\begin{center}
\begin{table*}[htbp]
\centering
\begin{tabular}{m{2.4cm}<{\centering}m{6cm}<{\centering}m{5cm}<{\centering}m{2.5cm}<{\centering}}
\hline
\hline
TNE quantities & definition & component & physical meaning \\
\hline
$\bm{\Delta}^{\sigma*}_{2}$ & $\sum_{i}(f_{i}^{\sigma}-f^{\sigma,eq}_{i})\mathbf{v}^{*}_{i}\mathbf{v}^{*}_{i}$ & $\Delta^{\sigma*}_{2,xx}$,$\Delta^{\sigma*}_{2,xy}$,$\Delta^{\sigma*}_{2,yy}$ & non-organized momentum flux (NOMF) \\
\hline
$\bm{\Delta}^{\sigma*}_{3,1}$ &
$\frac{1}{2}\sum_{i}(f_{i}^{\sigma}-f^{\sigma,eq}_{i})(\mathbf{v}^{*}_{i}\cdot\mathbf{v}^{*}_{i}+\eta_{i}^{\sigma2})\mathbf{v}^{*}_{i}$ & $\Delta^{\sigma*}_{3,1,x}$,$\Delta^{\sigma*}_{3,1,y}$ & non-organized energy flux (NOEF)  \\
\hline
$\bm{\Delta}^{\sigma*}_{3}$ & $\sum_{i}(f_{i}^{\sigma}-f^{\sigma,eq}_{i})\mathbf{v}^{*}_{i}\mathbf{v}^{*}_{i}\mathbf{v}^{*}_{i}$  & ${\Delta}^{\sigma*}_{3,xxx}$,${\Delta}^{\sigma*}_{3,xxy}$,${\Delta}^{\sigma*}_{3,xyy}$,${\Delta}^{\sigma*}_{3,yyy}$ & the flux of  $\bm{\Delta^{\sigma*}_{2}}$  \\
\hline
$\bm{\Delta}^{\sigma*}_{4,2}$ & $\frac{1}{2}\sum_{i}(f_{i}^{\sigma}-f^{\sigma,eq}_{i})(\mathbf{v}^{*}_{i}\cdot\mathbf{v}^{*}_{i}+\eta_{i}^{\sigma2})\mathbf{v}^{*}_{i}\mathbf{v}^{*}_{i}$ & ${\Delta}^{\sigma*}_{4,2,xx}$,${\Delta}^{\sigma*}_{4,2,xy}$,${\Delta}^{\sigma*}_{4,2,yy}$ & the flux of  $\bm{\Delta^{\sigma*}_{3,1}}$  \\
\hline
$D^{\sigma*}_{2}$ & $\sum_{ix,iy} \sqrt{(\Delta^{\sigma*}_{2,xx})^2+(2\Delta^{\sigma*}_{2,xy})^2+(\Delta^{\sigma*}_{2,yy})^2}$ & scalar & TNE strength from the $\Delta^{\sigma*}_{2}$ \\
\hline
$D^{\sigma*}_{3,1}$ & $\sum_{ix,iy} \sqrt{(\Delta^{\sigma*}_{3,1,x} )^2+( \Delta^{\sigma*}_{3,1,y} )^2}$ & scalar & TNE strength from the $\Delta^{\sigma*}_{3,1}$ \\
\hline
$D^{\sigma*}_{3}$ & $\sum_{ix,iy}\sqrt{(\Delta^{\sigma*}_{3,xxx} )^2+(2\Delta^{\sigma*}_{3,xxy} )^2+(2\Delta^{\sigma*}_{3,xyy} )^2+( \Delta^{\sigma*}_{3,yyy} )^2}$ & scalar & TNE strength from the $\Delta^{\sigma*}_{3}$ \\
\hline
$D^{\sigma*}_{4,2}$ & $\sum_{ix,iy}\sqrt{(\Delta^{\sigma*}_{4,2,xx})^2+(2\Delta^{\sigma*}_{4,2,xy})^2+(\Delta^{\sigma*}_{4,2,yy})^2}$ & scalar & TNE strength from the $\Delta^{\sigma*}_{4,2}$ \\
\hline
$D^{*}_{m}$ & $D^{A*}_{m}+D^{B*}_{m}$ & scalar & TNE strength from the $D^{*}_{m}$ \\
\hline
$D^{*}_{m,n}$ & $D^{A*}_{m,n}+D^{B*}_{m,n}$ & scalar & TNE strength from the $D^{*}_{m,n}$ \\
\hline
\hline
\end{tabular}
\caption{ The definitions and the corresponding physical meanings of the common TNE quantities in DBM, where the operator $\sum_{ix,iy}$ indicates integrating over all the fluid units and multiply the unit area $dx dy$.
From a certain perspective, the TNE strength is increasing;
While from a different perspective, the TNE strength, on the other hand, may be decreasing.
It is one of the concrete manifestations of the complexity of non-equilibrium flow behavior.
}
\label{Table2}
\end{table*}
\end{center}

\section{ Numerical simulations and results }\label{Numerical simulations}

In this section, we first validate the DBM code by comparing the DBM results with experimental results.
Then, the effects of specific-heat ratio on the dynamic process and TNE behaviors on SBI are investigated.

\subsection{ Comparison with experimental results } \label{experiment}

\begin{figure}[htbp]
\center\includegraphics*
[width=0.5\textwidth]{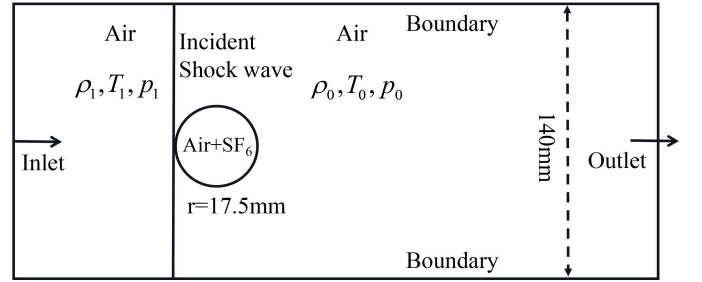}
\caption{ The computational configuration of the shock-bubble interaction.}
\label{fig2}
\end{figure}

In the following part, we use a first-order two-fluid DBM to simulate the interaction between a planar shock wave with a 2-D heavy-cylindrical bubbles, and compare the DBM results with the experimental results from Ref. \cite{Ding2018POF}.
The computational configuration can be seen in Fig. \ref{fig2}.
In a flow field which is filled with $\rm{Air}$, there is a static bubble composed of $26\%$ Air and $74\% \; \rm{SF_6}$.
A shock with $\rm{Ma}=1.2$ would pass through the bubble from left to right.
The initial conditions of ambient gas are $\rho_0=1.29\rm{kg/m^3}$, $T_0=293\rm{K}$, $p_0=101.3 \rm{kPa}$.
Ignoring the pressure difference between interior gas and ambient gas, the initial parameters of the bubble are $\rho_{\rm{bubble}}=4.859\rm{kg/m^3}$, $p_{\rm{bubble}}=101.3 \rm{kPa}$, and $T_0=293\rm{K}$.
For simulating, these actual physical quantities should be transferred to dimensionless parameters.
This process can refer to the \ref{sec:AppendixesA}.
The dimensionless conditions of macroscopic quantities of the fluid field in initial time are
\[
\left\{ \begin{gathered}
(\rho,T,u_x ,u_y )_{\rm{bubble}} = (4.0347,1.0,0.0,0.0), \hfill \\
(\rho,T,u_x ,u_y )_1 = (1.3416,1.128,0.3616,0.0), \hfill \\
(\rho,T,u_x ,u_y )_0 = (1.0,1.0,0.0,0.0), \hfill \\
\end{gathered} \right.
\]
where the subscript ``0'' (``1'') represents downstream (upstream) region.

In two-fluid DBM code, the distribution function $f^{\rm{Air}}$ is used to describe the ambient gas, i.e., Air.
The $f^{\rm{bubble}}$ characters the bubble which is a mixture that composed of Air and $\rm{SF_{6}}$.
The grid number is $N_x \times N_y = 800 \times 400$, where the $N_x$ and $N_y$ are grid number in $x$ and $y$ direction, respectively.
This grid size has passed the mesh convergence test.
The below results also show that it is sufficient to meet the requirements of the following research problem.
Other parameters used for the simulation are: $c=1.0$, $\eta_{\rm{Air}}=\eta_{\rm{bubble}}=10.0$, $I _{\rm{Air}}=3$, $I_{\rm{bubble}}=15$, $\Delta x = \Delta y = 1.2 \times 10^{-4}$ and $\Delta t = 1 \times 10^{-6}$.
The viscosity effect is feeble compared to the shock compression effect, so it does not significantly affect the deformation of the bubble.
Therefore, in this part, the relaxation time $\tau$ is set sufficiently small.
The inflow (outflow) boundary condition is used in the left (right) boundary, and the periodic boundary is adopted in the $y$ direction.
The first-order forward difference scheme is used to calculate the temporal derivative, and the second-order nonoscillatory nonfree dissipative scheme is adopted to solve the spatial derivative in Eq. (\ref{Eq.Discrete-Boltzmann1}) \cite{Lai2016PRE,Zhang2020POF,Zhang2022POF}.

\begin{figure*}[htbp]
\center\includegraphics*
[width=1.0\textwidth]{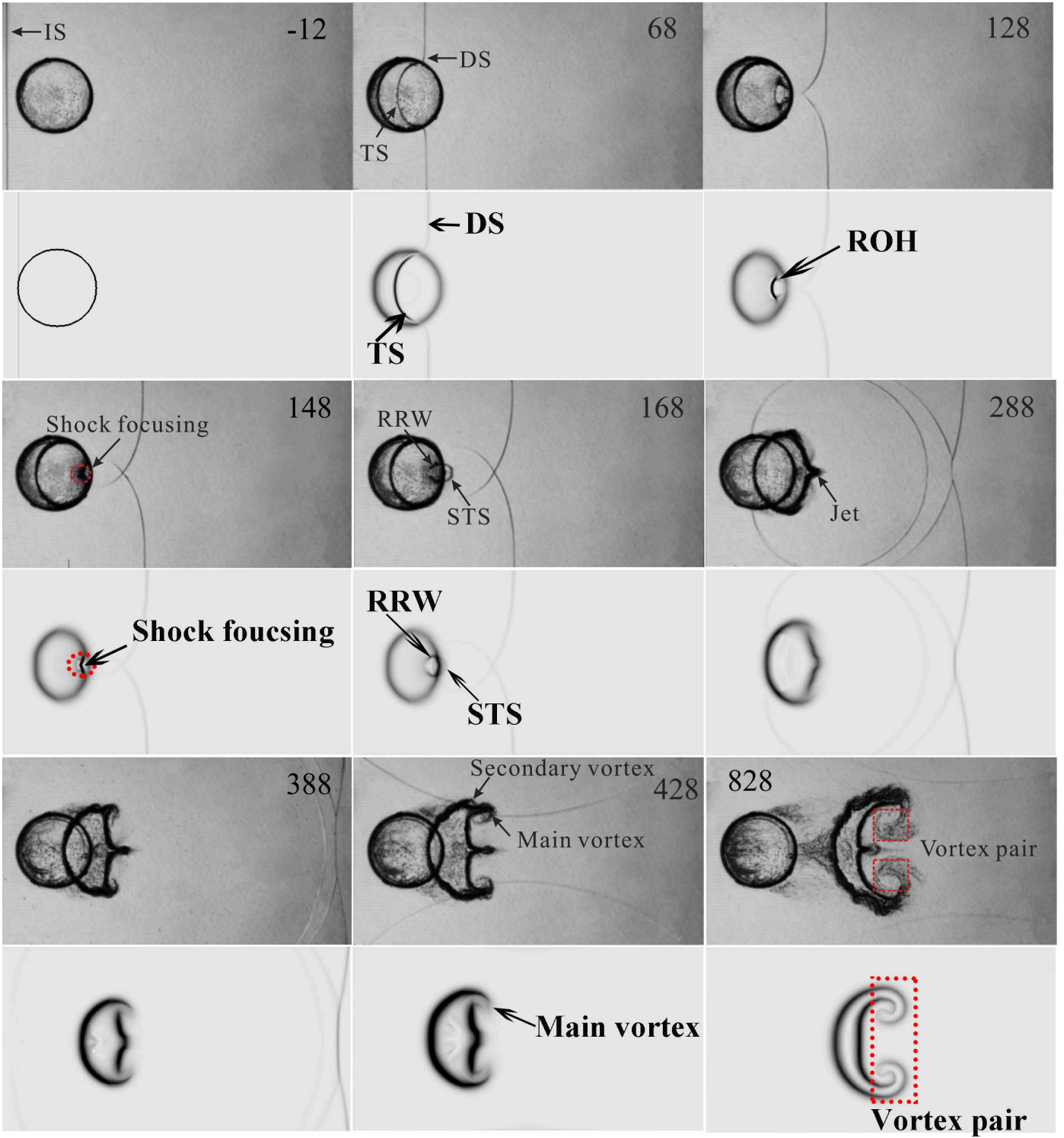}
\caption{  Snapshots of schlieren images of the interaction between a shock wave and a heavy-cylindrical bubble.
The odd rows represent experimental results from Ref. \cite{Ding2018POF}  with permission, and the even rows are DBM simulation results.
The typical wave patterns and the bubble's main characteristic structure are marked out in the figures.
Numbers in the picture represent the time in $\mu s$.
}
\label{fig3}
\end{figure*}

Two quantitative comparisons between experimental results and DBM simulations are shown in the following part, including snapshots of schlieren images and evolutions of characteristic scales for the bubble.
The first is shown in Fig. \ref{fig3}.
In the figure, results from odd rows are experimental, and the even rows indicate DBM simulation results.
The typical wave patterns and bubble's main characteristic structures are marked out in the figures.
Numbers in the pictures represent the time in $\mu s$.
Schlieren images of DBM results are calculated from the density gradient formula, i.e., $|\nabla \rho|/ |\nabla \rho|_{\rm{max}}$, with $|\nabla \rho|=\sqrt{(\partial \rho / \partial x)^2+(\partial \rho / \partial y)^2}$.
At $t=0 \mu s$, the incident shock wave impacts the upstream interface, and subsequently generates a transmitted shock (TS) propagating downstream in the bubble and a reflected shock wave moving upward in ambient gas.
The incident shock wave travels downstream continuously to form a diffracted shock (DS).
As TS propagates, it will split into three branches due to the considerable pressure perturbations caused by the gradual decay of the DS strength \cite{Wang2015POF}.
Afterward, as shown in the subfigure at about $t=128 \mu s$, two high pressure regions (ROH) generate because of the interaction of these branches.
Subsequently, at about $t=148 \mu s$, the two ROHs meet, causing the shock focusing.
On the one hand, at about $t=168 \mu s$, the shock focusing causes the generation of downstream-propagating second transmitted shock (STS) and upward-moving rarefaction wave.
On the other hand, it will produce high pressure region inside the bubble, which later leads to a jet structure, as shown at about $t=288 \mu s$.
At about $t=428 \mu s$, due to the deposited vorticity, there will produce a pair of counter-rotating vortexes at the pole region of the bubble.
The further development of the vortex pair and the effect of viscosity decrease the amplitude of the jet.
Finally, the jet structure disappears.

\begin{figure}[htbp]
\center\includegraphics*
[width=0.4\textwidth]{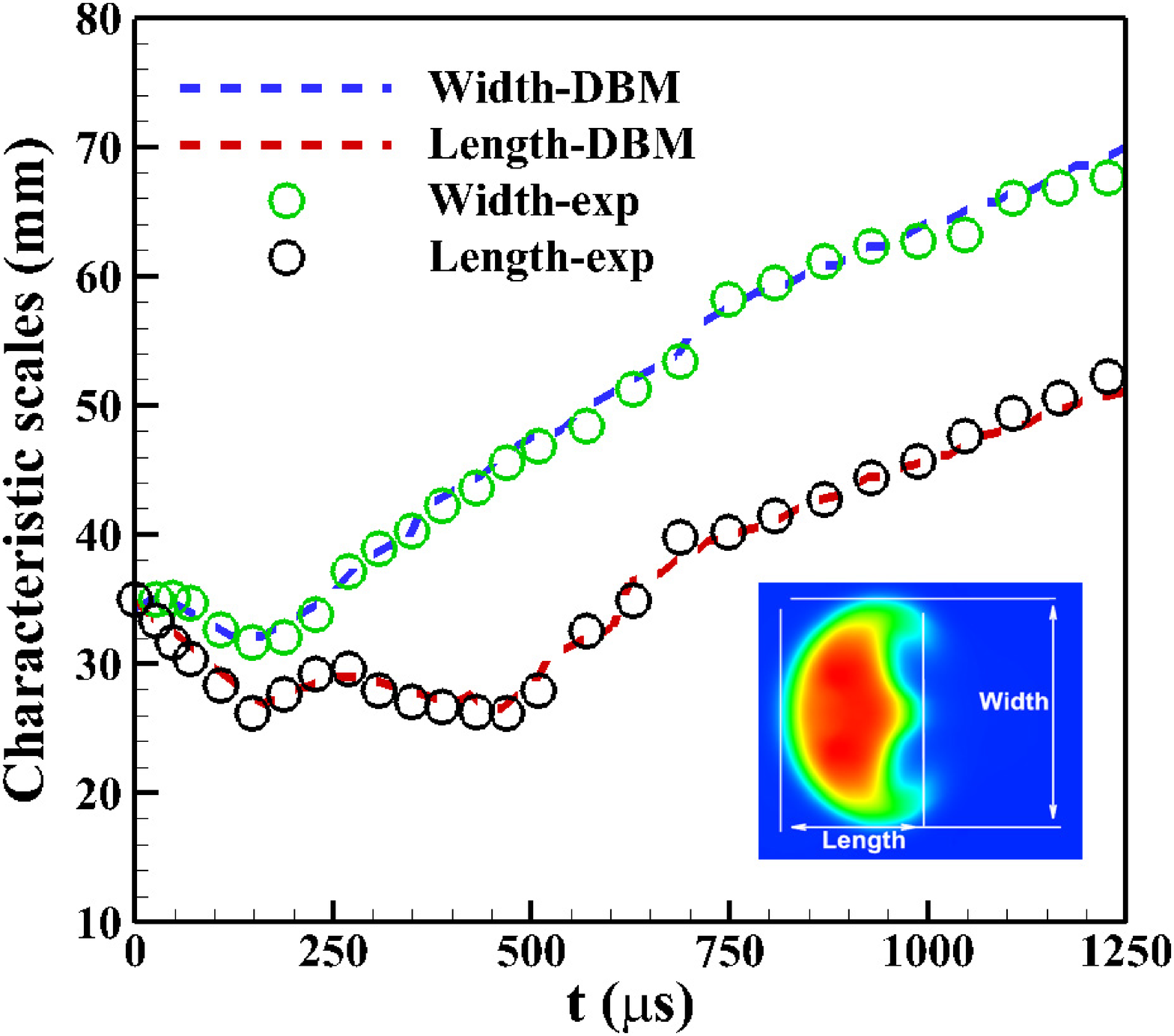}
\caption{ \textcolor{red}{The temporal variations of length and width of the bubble.
The symbols represent DBM results and the lines are experimental.
The definition of the length and the width of the bubble can be seen in the illustration.
Experimental results are obtained from Fig. 12, in Ref. \cite{Ding2018POF} with permission. }}
\label{fig4}
\end{figure}

The second quantitative comparison is the interface structure described by the length and width of the bubble, as shown in Fig. \ref{fig4}.
The experimental data are extracted from Fig. 12, in Ref. \cite{Ding2018POF}.
Quantitative agreements between DBM simulation and experimental results are seen.
For the profile of bubble width, there are mainly two stages.
At an early time ($t<150\mu s$), it decreases to a minimum value because of the shock compression effect.
After the shock wave passes through the bubble ($t>150\mu s$), the developed vortex pair caused by the deposited vorticity gradually dominates the growth of bubble width.
Different from width evolution, the temporal variation of length experiences three stages.
In the early stages ($t<150\mu s$), it decreases quickly due to the shock compression effect.
Then, the jet structure emerges, which results in a growth in length ($150\mu s<t<250\mu s$).
Because the upstream interface moves faster than the downstream interface, the bubble length would decrease at $250\mu s<t<500\mu s$.
In the third stage ($t>500\mu s$), the vortex pair forms and then leads to a continuous development of bubble length.
Both the length and width experience oscillations in the later stages due to complex wave patterns.

The quantitative agreements between DBM simulation and experimental results indicate the following two facts: (i) the order of TNE considered in the current DBM is sufficient, (ii) the choosing of discrete velocities and spatial-temporal steps and simulation parameters like the relaxation times is suitable for characterizing the deformation of bubble, wave patterns, main characteristics of flow morphology.

\subsection{ Effects of specific-heat ratio on SBI }

The major of current works on SBI research have not focused on specific-heat ratio effects.
In this part, the simulation parameters are fine-adjusted based on the parameters in Section \ref{experiment} to highlight the influence of specific-heat ratio.
Through adjusting the extra degree of freedom $I$, five cases with various specific-heat ratios of the bubble are simulated, i.e., $\gamma=1.4, 1.28, 1.18, 1.12$, and $1.09$.
Two kinds of analysis methods, including tracer particle method and two-fluid model, are used to characterize qualitatively the macroscopic behaviors such as the shape, deformation process, mixing degree, etc.
The related TNE behaviors are also studied.

\subsubsection{ Effects of specific-heat ratio on jet shape, deformation process, and average motion}

\begin{figure*}[htbp]
\center\includegraphics*
[width=1.0\textwidth]{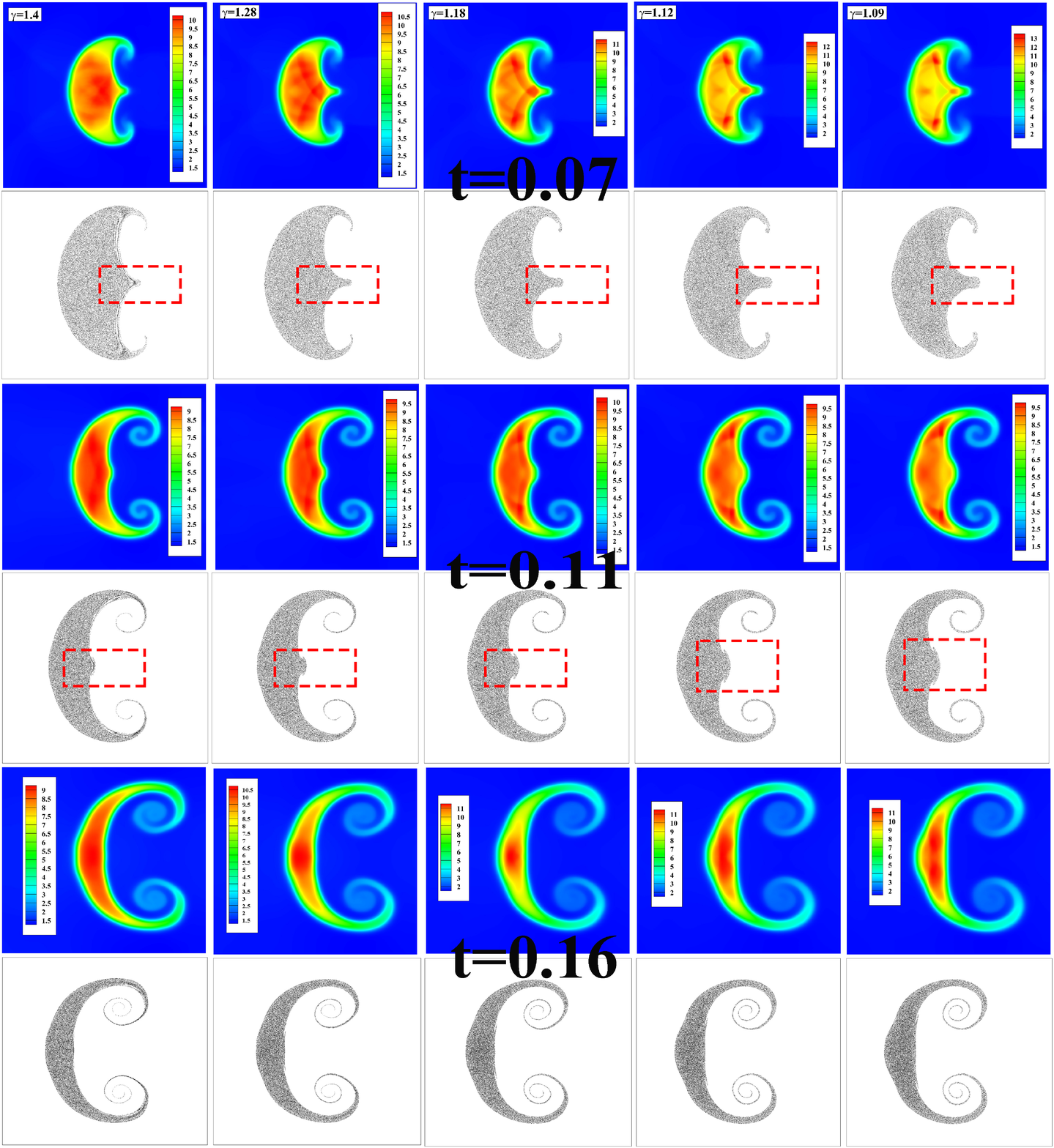}
\caption{ Density contours and particle tracer images at three different moments (i.e., $t=0.07, t=0.11$, and $t=0.16$) with various specific-heat ratios.
The odd rows represent density contours, and the even rows are tracer particle images. }
\label{fig8}
\end{figure*}

We first observe the specific-heat ratio effect on the bubble shape from the view of density contour and images of particle tracer visually.
As shown in Fig. \ref{fig8},
pictures with three typical moments are plotted, i.e., $t=0.07, t=0.11$, and $t=0.16$.
The odd rows represent density contours and the even rows are tracer particle images.
It can be seen that the specific-heat ratio significantly affects the length and shape of the jet structure.
The smaller the specific-heat ratio is, the stouter the jet structure can be seen.
The reason is that the specific-heat ratio significantly changes the propagation speed of shock waves and wave patterns inside the bubble.
The specific-heat ratio also influences the vortex structure in early stage but contributes little effects to it in later stage.
In the later stage, for cases with different specific-heat ratios, the differences in vortex pairs are almost invisible.

\begin{figure}[htbp]
\center\includegraphics*
[width=0.4\textwidth]{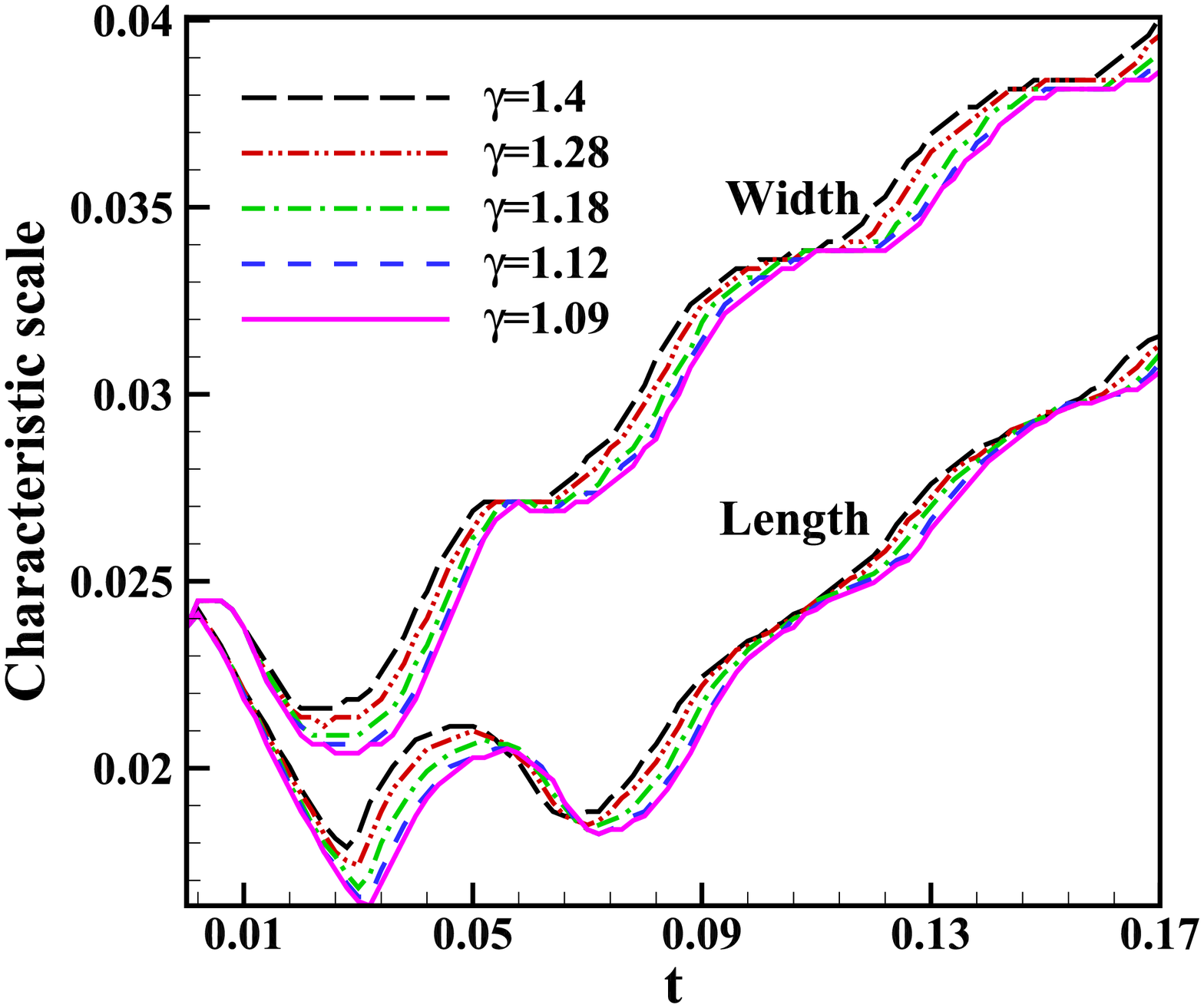}
\caption{ The temporal evolution of characteristic scales on SBI process, with different specific-heat ratios.
Lines with different colors represent the cases with various specific-heat ratios. }
\label{fig13}
\end{figure}

Then, the effects of specific-heat ratio on deformation process are analyzed.
Shown in Fig. \ref{fig13} are the evolutions of characteristic scales which used to describe the bubble size, i.e., width and length.
It can be seen that the smaller the specific-heat ratio of bubble, the smaller the bubble width and length.
For the fluid with smaller specific-heat ratio, it is easier to be compressed.
Therefore, the characteristic scales of bubbles with smaller specific-heat ratio tend to be compressed smaller.
It can also be seen that the case with the largest specific-heat ratio reaches the minimum characteristic scales firstly.
The reason is that the shock wave propagates faster in case with larger specific-heat ratio.

\begin{figure}[htbp]
\center\includegraphics*
[width=0.4\textwidth]{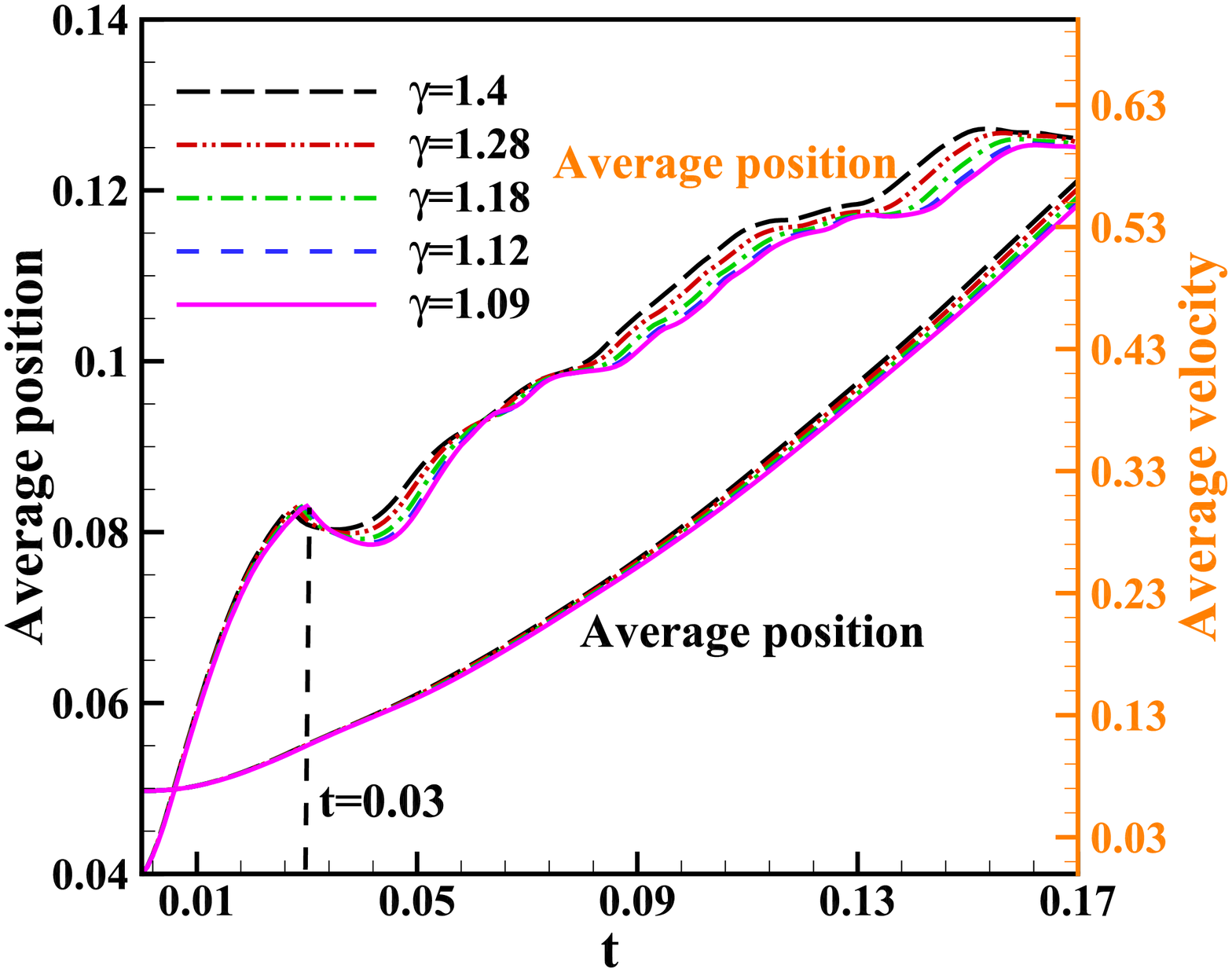}
\caption{ The temporal evolution of average position and average bubble velocity, with different specific-heat ratios.
Lines with different colors represent the cases with various specific-heat ratios. }
\label{fig14}
\end{figure}

Through the method of tracer, information on the average motion of the bubble is easy to obtain.
Shown in Fig. \ref{fig14} are the average position and average velocity of the bubble, with different specific-heat ratios.
It is found that, in the shock compression stage ($t<0.03$), the effect of specific-heat ratio contributes little to the average motion.
However, when the shock wave passes through the bubble ($t>0.03$), a larger specific-heat ratio speeds up the average motion of the bubbles.
The reason is that the bubbles with smaller specific-heat ratio need more energy to compress their size, so their translational energy is smaller.

\subsubsection{ Effects of specific-heat ratio on vortex motion}

\begin{figure*}[htbp]
\center\includegraphics*
[width=1.0\textwidth]{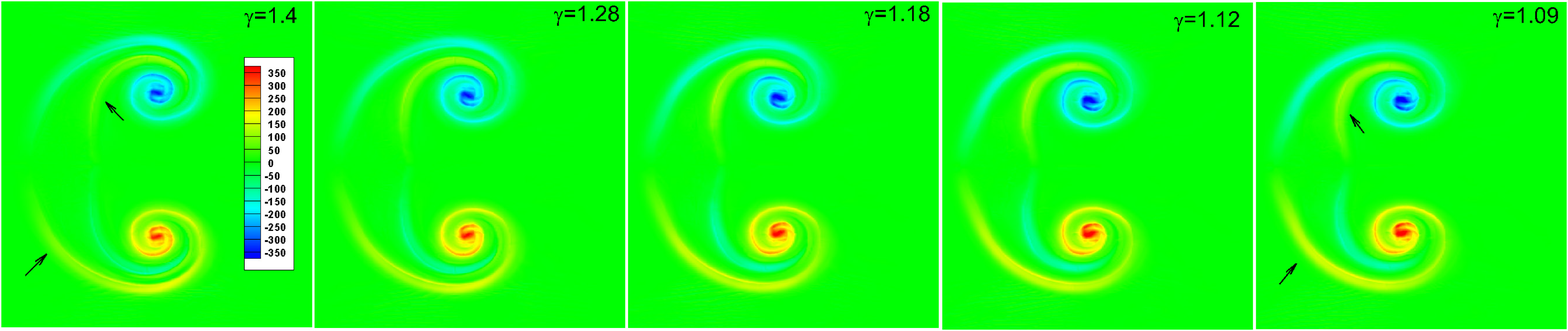}
\caption{ Vorticity contours at $t=0.134$, with various specific-heat ratios.
The arrows in the vorticity image point out the apparent difference between case $\gamma=1.4$ and case $\gamma=1.09$.
  }
\label{fig10}
\end{figure*}

Vorticity is one of the most important physical quantities in describing the vortex motion.
In the 2-D case, the vorticity can be calculated by the following equation:
\begin{equation}
\bm{\omega}=(\frac{\partial u_y}{\partial x} - \frac{\partial u_x}{\partial y}) \bm{e}_{z}
.
\end{equation}
The positive (negative) value of $\omega$ represents the positive (negative) direction along the $z$ axis.
Vorticity contours at $t=0.134$, with various specific-heat ratios, are shown in Fig. \ref{fig10}.
The discernable difference between cases with various specific-heat ratios can be observed.
The arrows in the vorticity images point out the obvious differences around the interface between case $\gamma=1.4$ and case $\gamma=1.09$.
That is to say, there exists influences of specific-heat ratio on the rotational motion of the bubble.

\begin{figure}[htbp]
\center\includegraphics*
[width=0.4\textwidth]{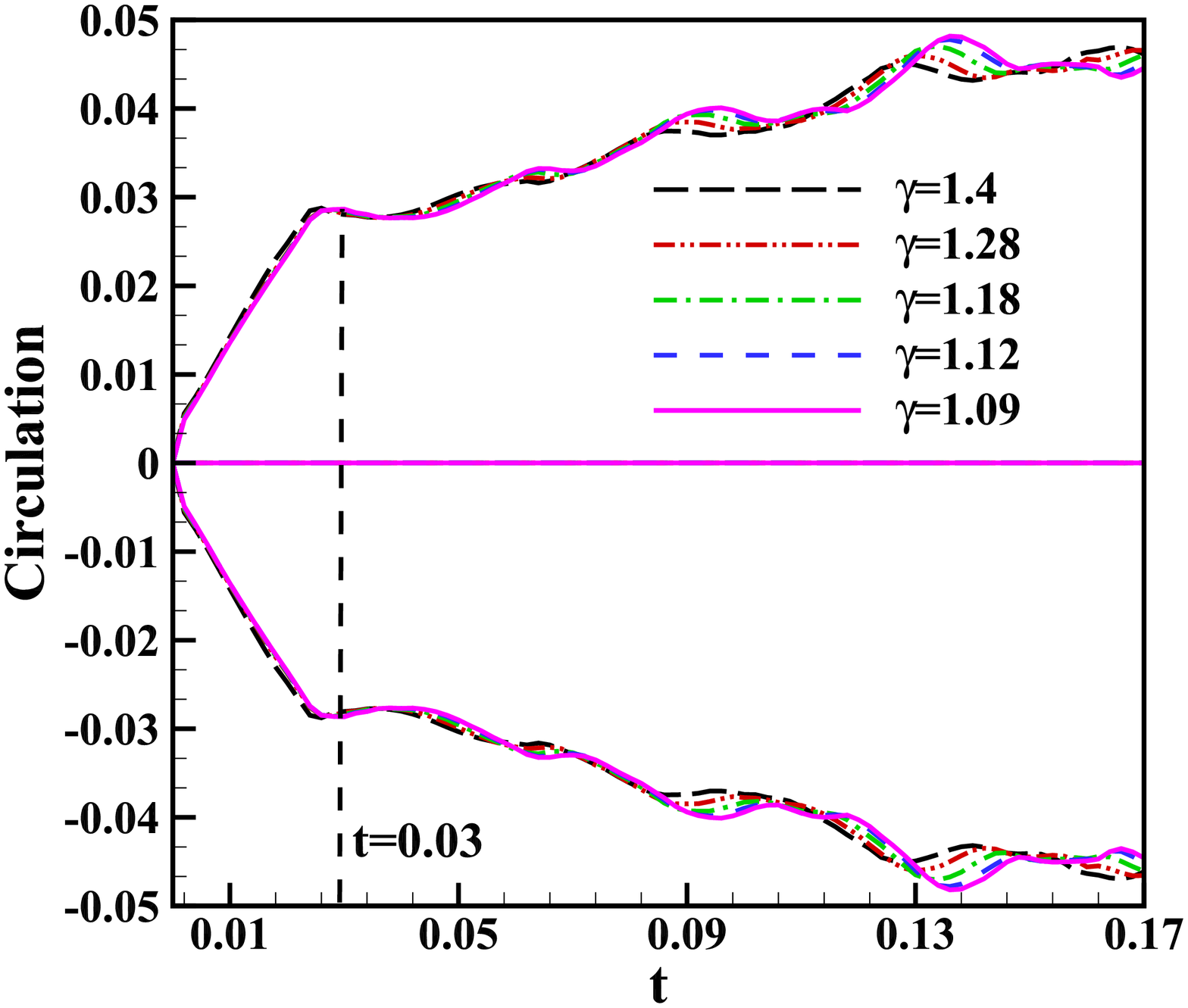}
\caption{ Temporal evolution of circulation on SBI process with various specific-heat ratios.
Lines with different colors represent the cases with various specific-heat ratios. }
\label{fig9}
\end{figure}

The strength of vorticity is described by circulation $\Gamma$, where $\Gamma=\sum \omega\Delta x\Delta y$.
$\Gamma^{+}=\sum \omega |_{\omega > 0} \Delta x\Delta y$ is the positive circulation and $\Gamma^{-}=\sum \omega |_{\omega < 0} \Delta x\Delta y$ represents the negative circulation.
Figure \ref{fig9} shows the temporal evolution of circulations on SBI process.
It can be seen that the values of $\Gamma$ are equal to zero all the time because the values of $\Gamma^{+}$ and $\Gamma^{-}$ are the same.
But they point in the opposite direction.
In the shock compression stage ($t<0.03$), the specific-heat ratio effect contributes little to the circulation of the bubble.
When the shock wave sweeps through the bubble ($t>0.03$), the specific-heat ratio affects the value of circulation obviously.
The cases with a smaller specific-heat ratio experiences a larger range of amplitude of change, which is caused by its good compressibility.

\subsubsection{ Effects of specific-heat ratio on mixing degree }

\begin{figure*}[htbp]
\center\includegraphics*
[width=1.0\textwidth]{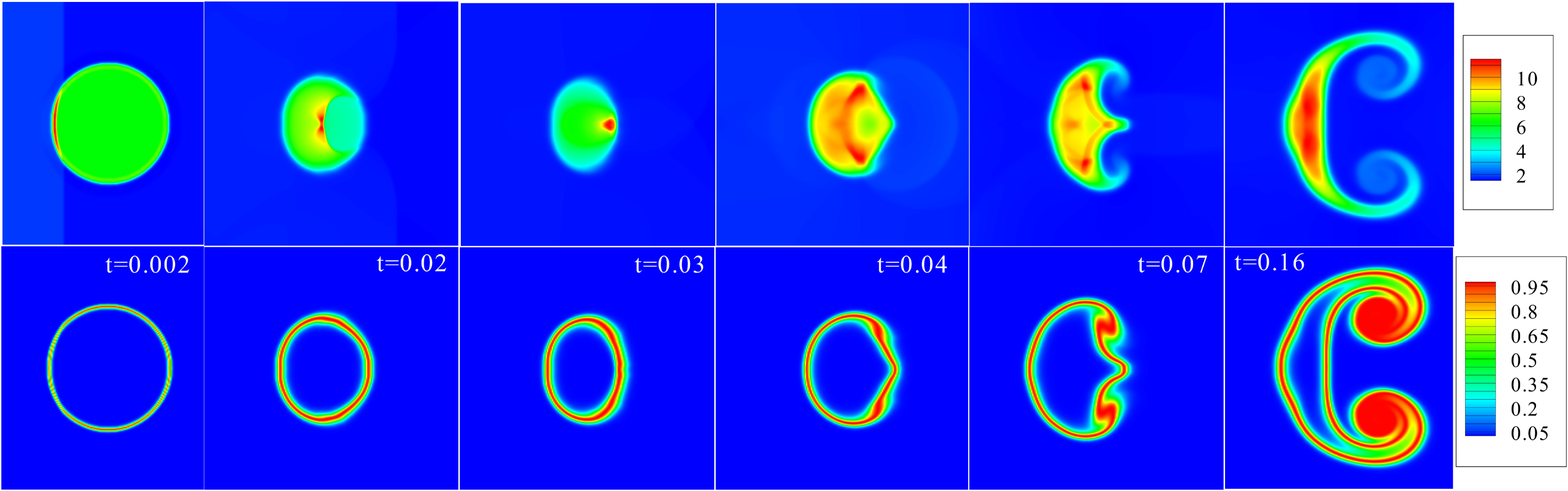}
\caption{ Density contours (first row) and mixing degree $M$ (second row) at several typical moments.  }
\label{fig6}
\end{figure*}

The mixing process is a fundamental research content on SBI.
In two-fluid DBM, the mixing degree at each fluid unit can be defined as $M=4 \cdot M_{A}M_{B}$, where $M_{\sigma}$ represents the mass fraction of component $\sigma$.
The higher the value of $M$, the higher the mixing amplitude.
Images of density (first cow) and mixing degree $M$ (second row) at several typical moments are shown in Fig. \ref{fig6}.
As can be seen, the mass mixing occurs in the region where two media contact.

\begin{figure}[htbp]
\center\includegraphics*
[width=0.4\textwidth]{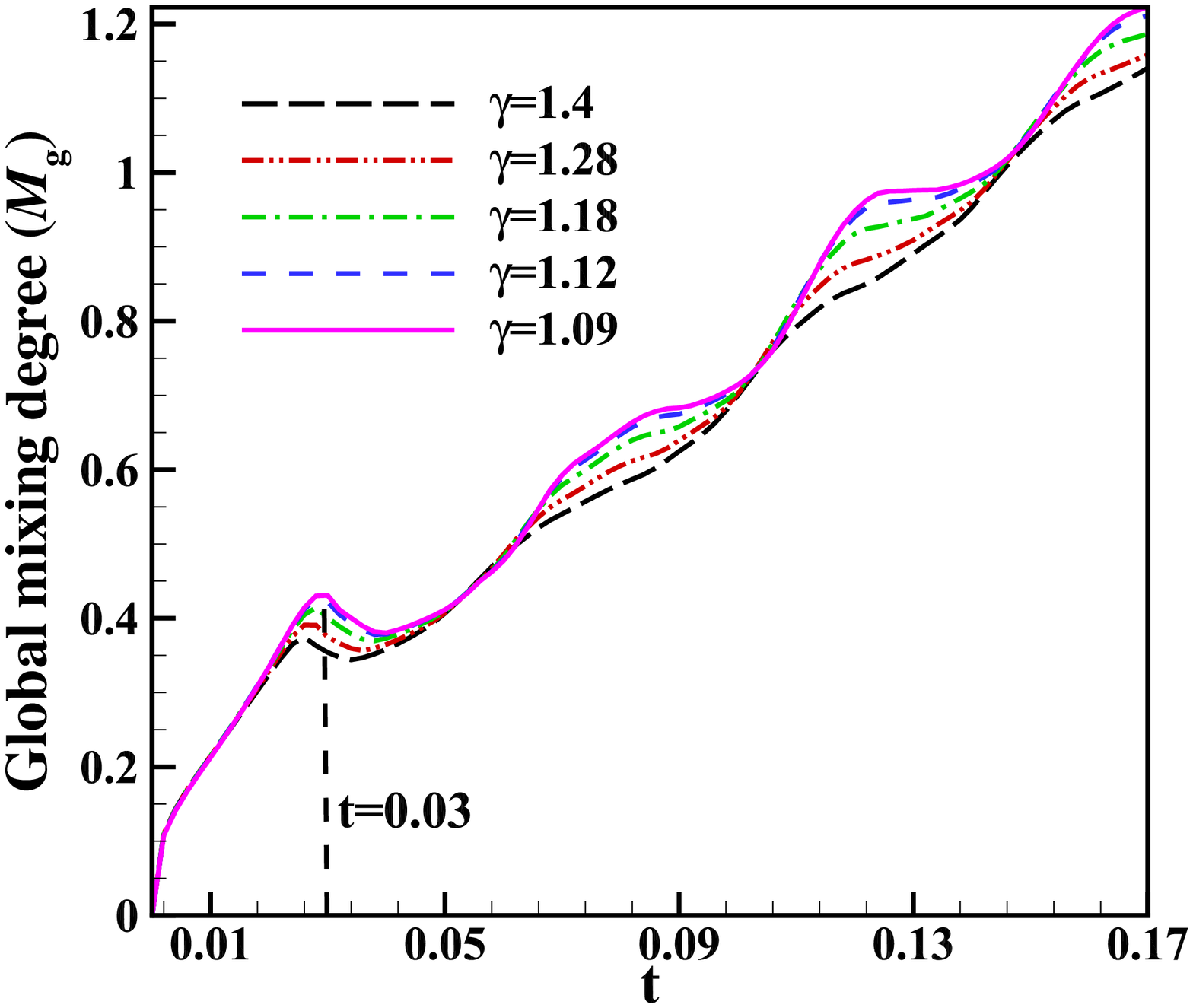}
\caption{ Temporal evolution of global mixing degree $M_{g}$ of SBI process with various specific-heat ratios.
Lines with different colors represent the cases with various specific-heat ratios. }
\label{fig5}
\end{figure}

In addition, the mixing degree $M_{g}$ described the whole fluid field can be defined
\begin{equation}
M_{g}=4 \cdot \frac{\overline{M_{A}\cdot M_{B}}}{\overline{M_{A}} \cdot \overline{M_{B}}}
,
\end{equation}
where  the symbol ``$-$'' indicates integrating the $M_{\sigma}$ over the whole fluid field and then dividing the grid size $N_x \cdot N_y$.
Shown in Fig. \ref{fig5} is the temporal evolution of the global mixing degree $M_{g}$.
As can be seen, temporal profiles of the global mixing degree show two stages: $t < 0.03$ and $t > 0.03$.
When $t < 0.03$, there is almost no difference between cases with various specific-heat ratios.
But for $t > 0.03$, the stronger the specific-heat ratio effect, the larger the mixing degree.
Actually, there are mainly two indicators that measure the global mixing degree: the amplitude of mixing and the area of the mixing zone between two fluids.
At the stage $t > 0.03$, the shock compression dominates the mix by enhancing the mixing amplitude and increasing the area of the mixing zone simultaneously.
In this stage, the specific-heat ratio effect contributes little to the mix.
However, when the shock passes through the bubble, the deformation of interface and the evolution of vortex core both significantly raise the area of the mixing zone.
As can be seen in Fig. \ref{fig5}, the smaller specific-heat ratio of bubble, the stronger global mixing degree of fluid field.
Intuitively, for the fluid with smaller specific-heat ratio, it is easier to deform and compress, which is beneficial for the fluid mixing.
It can also be explained by the diffusion formula, i.e., $J^{\sigma}_{\alpha}=-D^{\sigma}_{d}\frac{\partial \rho^{\sigma}}{\partial x}$, where $D^{\sigma}_{d}=\tau^{\sigma}T/m^{\sigma}$ is the diffusivity\cite{Zhang2020POF}.
The specific-heat ratio affects both the temperature $T$ and the gradient of density simultaneously.
Therefore, these two aspects comprehensively influence the material diffusion between the two fluids.
Due to the complex reflected shock wave, the global mixing degree shows a tendency for oscillating growth.

\subsubsection{ Effects of specific-heat ratio on TNE behaviors }
\begin{figure}[htbp]
\centering
\subfigure[]{
\begin{minipage}{8cm}
\centering
	\includegraphics[width=7cm]{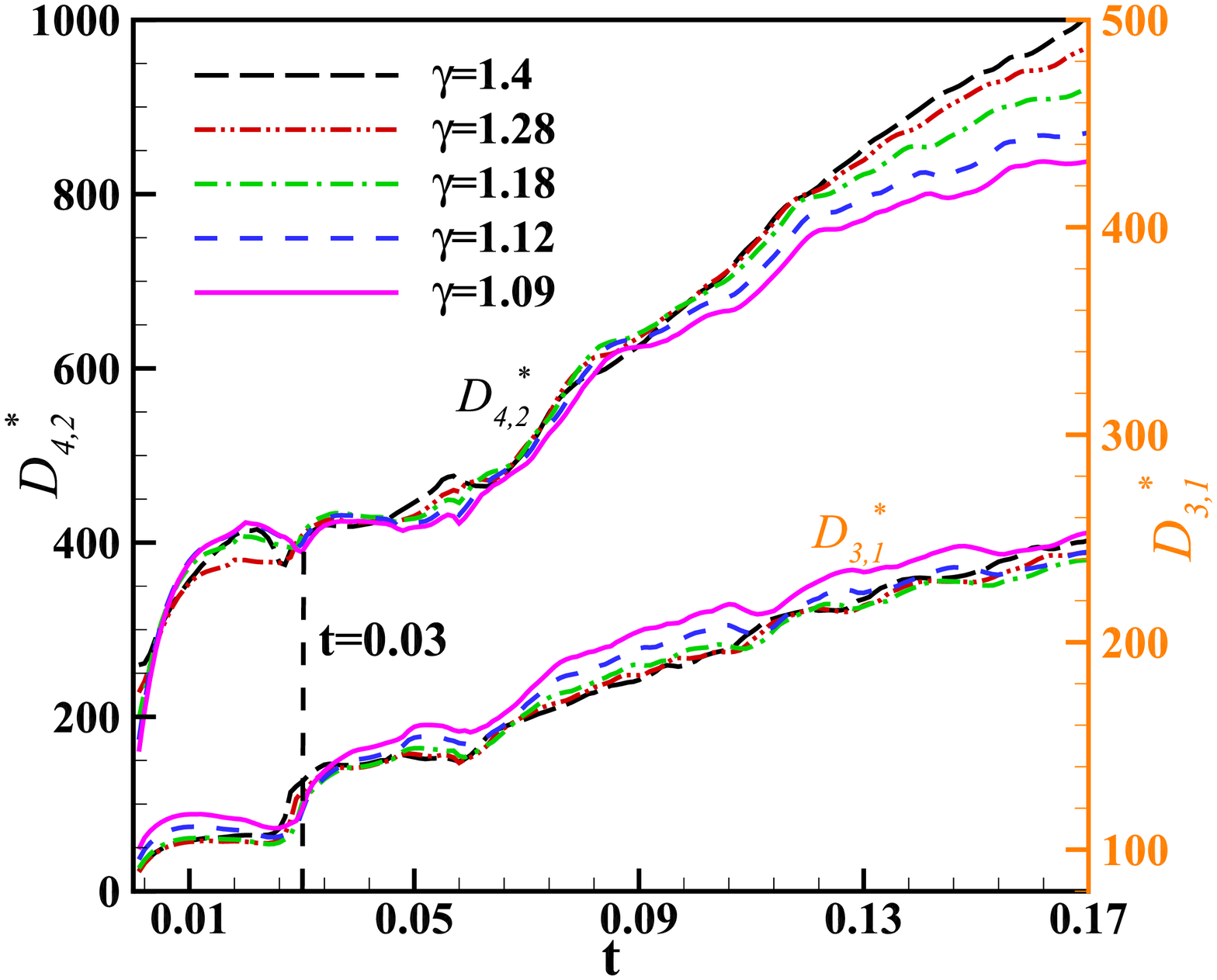}
	\label{fig12b}
\end{minipage}
}
\subfigure[]{
\begin{minipage}{8cm}
\centering
	\includegraphics[width=7cm]{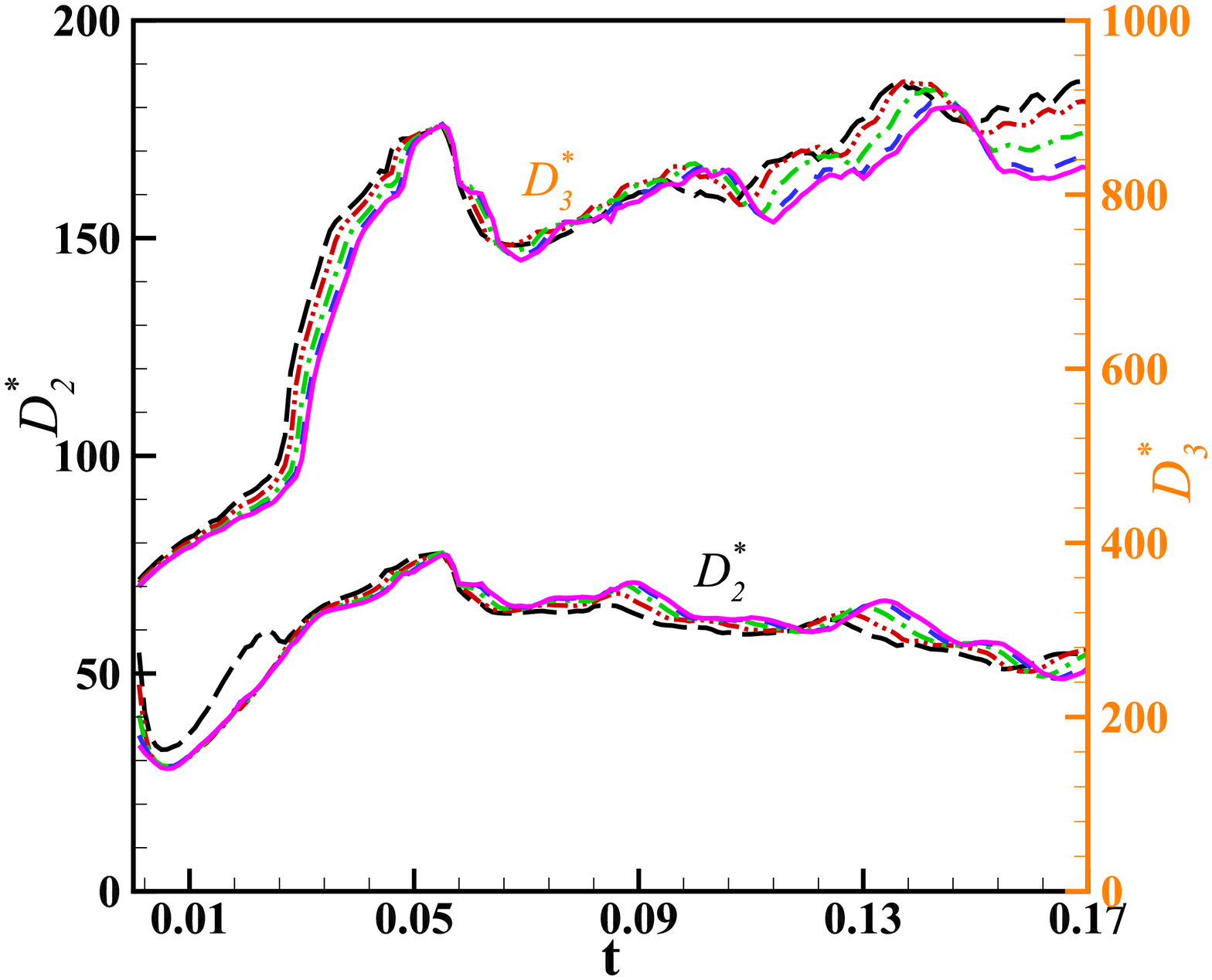}
	\label{fig12a}
\end{minipage}
}
\caption{ (a) Temporal evolution of $D_{3,1}^{*}$ and $D_{4,2}^{*}$.
(b) Temporal evolution of $D_{2}^{*}$ and $D_{3}^{*}$. Lines with different colors represent the cases with various specific-heat ratios. }
\label{fig12}
\end{figure}

The investigation of TNE behaviors is of great importance for understanding the kinetics process on SBI.
These TNE quantities describe the fluid system deviating from the thermodynamic state from their own perspectives.
The effects of specific-heat ratio on global TNE strength, i.e., $D_{2}^{*}$, $D_{3}^{*}$, $D_{3,1}^{*}$, and $D_{4,2}^{*}$, are shown in Fig. \ref{fig12}.
It can be seen that the effects of specific-heat ratios on various TNE quantities are different.
Theoretically, the influence of specific-heat ratio on the non-equilibrium effect is reflected in two aspects: transport coefficient and macroscopic quantity gradient.
For example, on the one hand, the specific-heat ratio reduces heat conductivity, while on the other hand, it enhances the temperature gradient.
Therefore, the effect of specific heat ratio on NOEF is the comprehensive result of the competition between the two aspects.
As shown in Fig. \ref{fig12b}, the smaller the specific-heat ratio, the stronger strength of $D_{3,1}^{*}$.
It indicates that the specific-heat ratio increase the strength of $D_{3,1}^{*}$ by raising the heat conductivity
\footnote{The heat conductivity formula is $\kappa=C_{p}\tau p$, with $C_{p}=\frac{D+I+2}{2}R$.
The larger the extra degree of freedom $I$ (the smaller the specific-heat ratio $\gamma$), the larger the heat conductivity $\kappa$.}.
For the strength of $D_{3}^{*}$, as shown in Fig. \ref{fig12a}, it is seen that it decreases as the specific-heat ratio becomes small.
The reason is that a smaller specific-heat ratio decreases the temperature gradient.
Effects of specific-heat ratio on $D_{4,2}^{*}$ show two-stage.
In the shock compression stage ($t<0.03$), the smaller specific-heat ratio, the larger the strength of $D_{4,2}^{*}$.
But the situation is reversed at the stage $t>0.03$.
For strength of $D_{2}^{*}$, the specific-heat effects are more significant in later stage.

\subsubsection{ Effects of specific-heat ratio on entropy production rate and entropy production }

The concepts of entropy are commonly used in complex flows \cite{Zhang2019Matter,Chen2022PRE,Sun2016IJHMT,Sun2016IJHMT2}.
In DBM, there are two kinds of entropy production rates, i.e., $\dot{S}_{\rm{NOEF}}$ and $\dot{S}_{\rm{NOMF}}$ \cite{Zhang2019Matter}.
They are key factors in compression science field.
The former is induced by temperature gradient and the NOEF ($\bm{\Delta_{3,1}^{*}}$). The latter is affected by velocity gradient and the NOMF ($\bm{\Delta_{2}^{*}}$).
The entropy production rates are defined by the following formulas \cite{Zhang2019Matter}:

\begin{equation}
\dot{S}_{\rm{NOEF}} =  \int \bm{\Delta_{3,1}^{*}} \cdot \nabla \frac{1}{T} d \bm{r}
,
\end{equation}

\begin{equation}
\dot{S}_{\rm{NOMF}} =  \int -\frac{1}{T} \bm{\Delta_{2}^{*}} : \nabla \bm{u} d \bm{r}
.
\end{equation}
Integrating the $\dot{S}_{\rm{NOEF}}$ and $\dot{S}_{\rm{NOMF}}$ over time $t$, the entropy generations over this period of time are obtained, i.e., $S_{\rm{NOEF}}=\int_{0}^{t} \dot{S}_{\rm{NOEF}} dt$ and $S_{\rm{NOMF}}=\int_{0}^{t} \dot{S}_{\rm{NOMF}} dt$.

\begin{figure}[htbp]
\centering
\subfigure[]{
\begin{minipage}{8cm}
\centering
	\includegraphics[width=7cm]{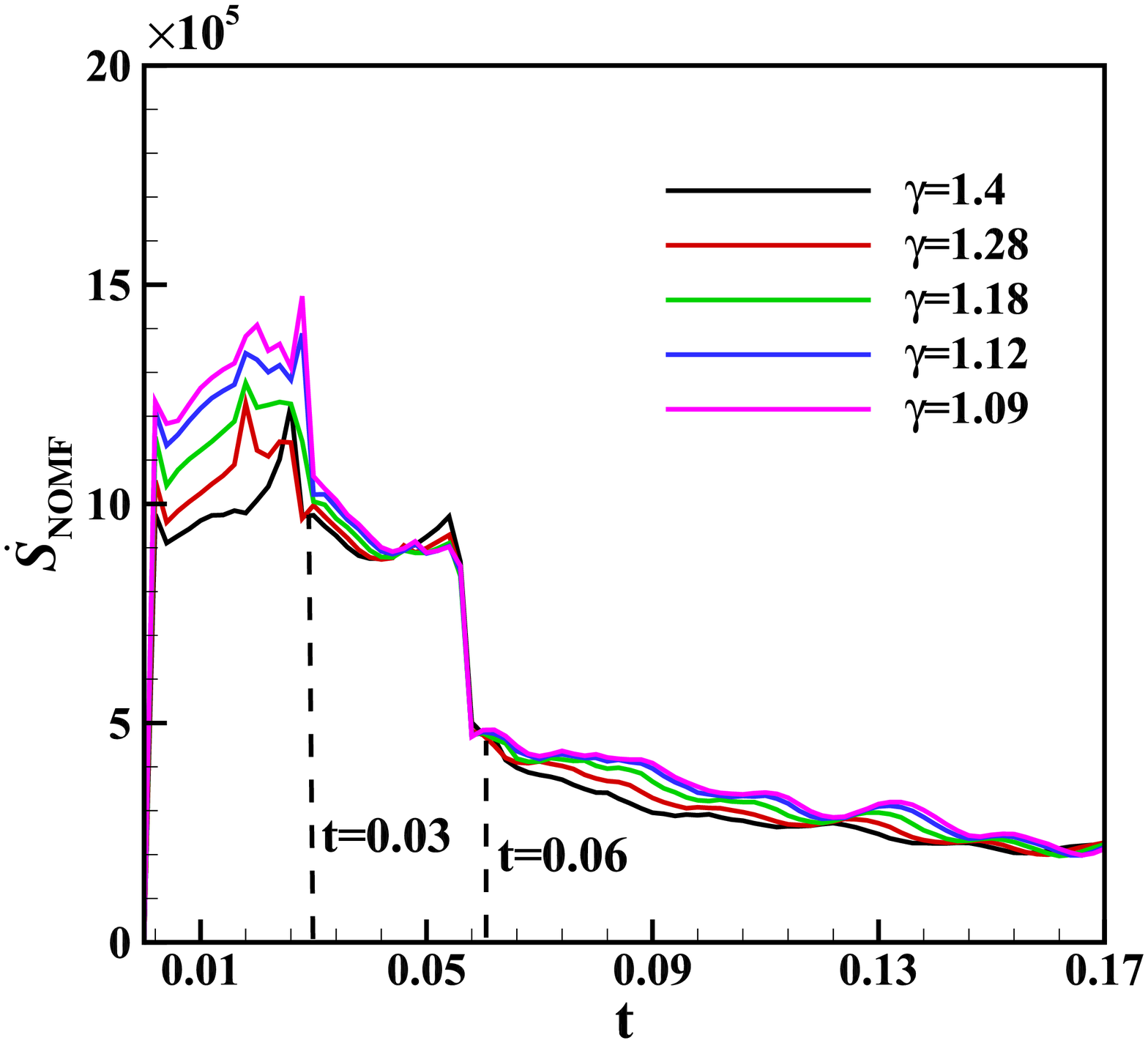}
	\label{fig7a}
\end{minipage}
}
\subfigure[]{
\begin{minipage}{8cm}
\centering
	\includegraphics[width=7cm]{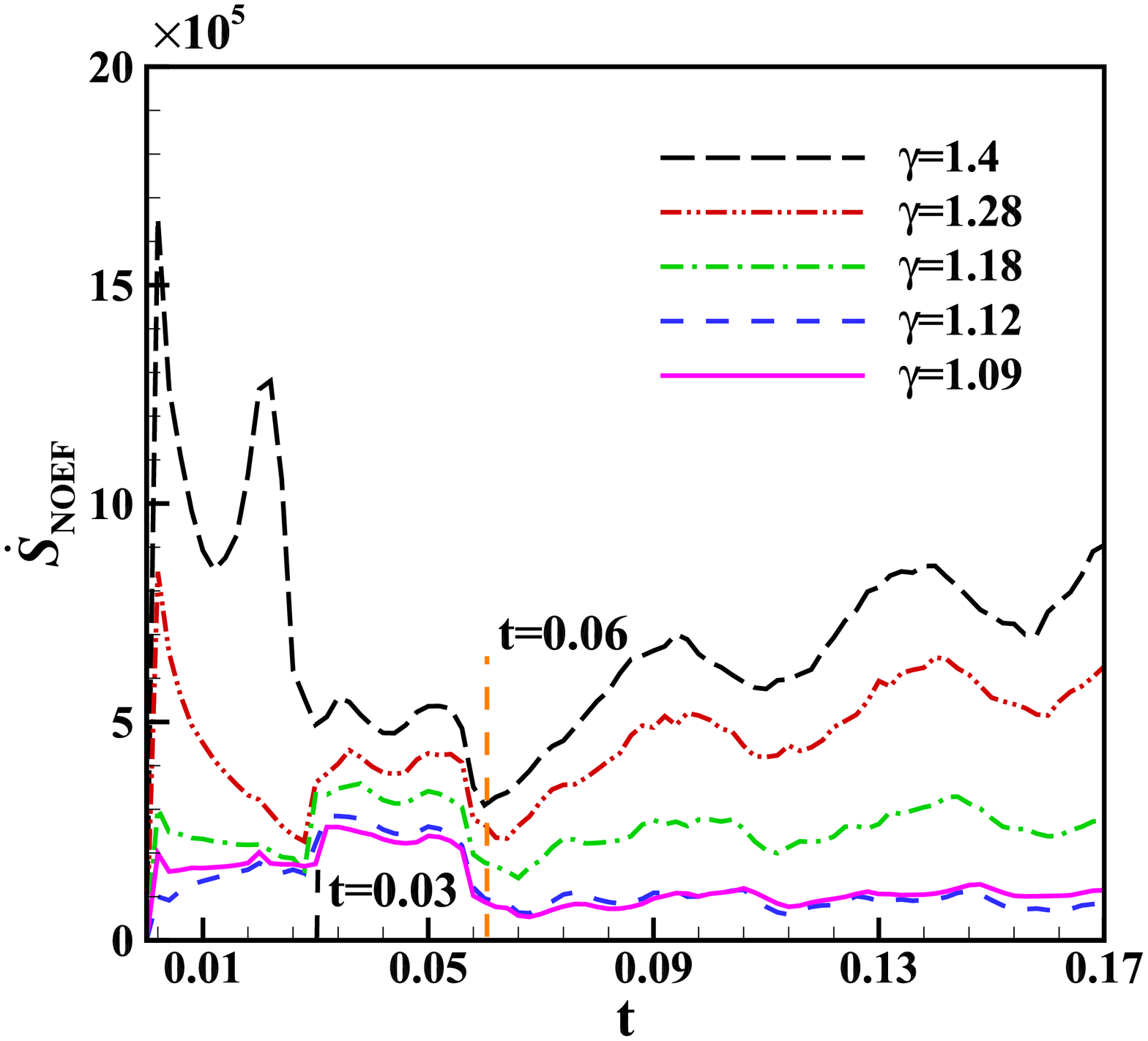}
	\label{fig7b}
\end{minipage}
}
\caption{ (a) Temporal evolution of entropy production rate $\dot{S}_{\rm{NOMF}}$.
(b) Temporal evolution of entropy production rate $\dot{S}_{\rm{NOEF}}$.
Lines with different colors represent the cases with various specific-heat ratios.  }
\label{fig7}
\end{figure}

Plotted in Fig. \ref{fig7}(a) and \ref{fig7}(b) are the temporal evolution of $\dot{S}_{\rm{NOMF}}$ and $\dot{S}_{\rm{NOEF}}$, respectively.
The evolution of entropy generation rate is related to two aspects: (i) the propagation of the shock wave, and (ii) the deformation of the bubble.
The former generates a macroscopic quantity gradient, and the latter makes the contact interface wider, longer, and deformed.
Depending on the location of the shock wavefront, there exist two critical moments in this SBI process: (i) at around $t=0.03$, the shock wave just sweeps through the bubble, and (ii) at $t=0.06$, the shock wave exits the flow field.
Therefore, the temporal evolution of the entropy production rate shows three stages, i.e., $t<0.03$, $0.03<t<0.06$, and $t>0.06$.
At the stage $t<0.03$, the shock compression stage, the shock effects compress the bubble.
It generates the large macroscopic quantity gradients, resulting in a quick increase of $\dot{S}_{\rm{NOMF}}$.
At around $t=0.03$, the shock wave passed through the bubble.
So the values of $\dot{S}_{\rm{NOMF}}$ decreases.
The values of $\dot{S}_{\rm{NOMF}}$ would continue to decrease due to the gradually wider contact interface caused by the diffusion effect.
At around $t=0.06$, the shock wave comes out of the flow field so that the values of $\dot{S}_{\rm{NOMF}}$ drops rapidly.
In the third stage, i.e., $t>0.06$, because of the diffusive effect, the general trend of $\dot{S}_{\rm{NOMF}}$ is downward.
However, it shows an oscillatory trend due to the influence of various reflected shock waves.
The specific heat ratio indirectly changes the value of $\dot{S}_{\rm{NOMF}}$ by changing the velocity gradient.
The smaller the specific-heat ratio, the larger $\dot{S}_{\rm{NOMF}}$.

Different understanding can be seen in Fig. \ref{fig7}(b), where the temporal evolution of $\dot{S}_{\rm{NOEF}}$ is plotted.
In the first stage ($t<0.03$), cases with different specific-heat ratios show various trends.
At the stage where the bubble deformation is not very large, i.e., $0.03<t<0.06$, values of $\dot{S}_{\rm{NOEF}}$ fluctuate near the average value.
In the third stage ($t>0.06$), evolutions of $\dot{S}_{\rm{NOEF}}$ in cases with larger specific-heat ratios show an apparent growing tendency.
Differently, the values of $\dot{S}_{\rm{NOEF}}$ in cases with smaller specific-heat ratios remain almost unchanged.
The influence of specific heat ratio on the $\dot{S}_{\rm{NOEF}}$, similar with the effect on  NOEF, is also affected by the heat conductivity and the temperature gradient.
It can be seen that, except for the case of $\gamma=1.09$, the larger the specific-heat ratio, the higher entropy production rate $\dot{S}_{\rm{NOEF}}$.
The temporal evolutions of $\dot{S}_{\rm{NOEF}}$ of case $\gamma=1.09$ and case $\gamma=1.12$ are very similar.
Consequently, the specific-heat ratio increases the $\dot{S}_{\rm{NOEF}}$ by raising the temperature gradient.

\begin{figure}[htbp]
\center\includegraphics*
[width=0.4\textwidth]{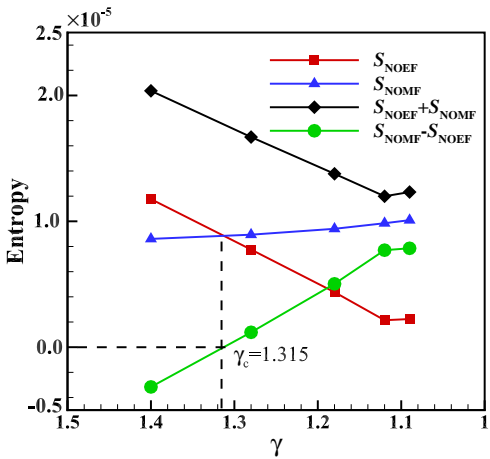}
\caption{ Entropy production ($S_{\rm{NOMF}}$, $S_{\rm{NOEF}}$, $S_{\rm{NOMF}}$+$S_{\rm{NOEF}}$, and $S_{\rm{NOMF}}$-$S_{\rm{NOEF}}$) over this period.  }
\label{fig11}
\end{figure}

Further understanding can be seen in Fig. \ref{fig11}, where the entropy productions over this period are plotted.
For convenience, the sum and difference between $S_{\rm{NOMF}}$ and $S_{\rm{NOEF}}$ are also plotted in the figure.
The variation range of $S_{\rm{NOEF}}$ is larger than that of $S_{\rm{NOMF}}$.
It indicates that the influence of specific-heat ratio on $S_{\rm{NOEF}}$ is more significant than that on $S_{\rm{NOMF}}$.
Effects of specific-heat ratio on entropy production caused by NOMF and NOEF are contrary.
Specifically, it can be seen that the entropy production contributed by NOMF increases with reduced specific-heat ratio.
But the entropy production caused by NOEF first reduces with decreasing specific-heat ratio and then approaches to a saturation value.
The $S_{\rm{NOEF}}$ in case $\gamma=1.09$ is almost the same with it in case $\gamma=1.12$.
When the specific-heat ratio $\gamma$ is smaller than a threshold value $\gamma_c$ ($\gamma_c \approx 1.315$), the entropy production induced by NOEF is more significant than that caused by NOMF.
However, in the case of $\gamma>\gamma_c$, the situation reverses.
The temporal evolution of the total entropy production ($S_{\rm{NOMF}}$+$S_{\rm{NOEF}}$) is similar to the $S_{\rm{NOEF}}$ profile.
The difference between $S_{\rm{NOMF}}$ and $S_{\rm{NOEF}}$ increases with decreasing specific-heat ratio.

\section{Conclusions}\label{Conclusions}

Specific-heat ratio effects on the interaction between a planar shock wave and a 2-D heavy-cylindrical bubble are studied by a two-fluid DBM which has a  flexible specific-heat ratio and  includes several schemes for analyzing the complex physical fields.
Besides the HNE that NS easily describes, the DBM pays more attention to the related TNE that NS is not convenient to describe.
First, both the snapshots of schlieren images and evolutions of characteristic scales from DBM simulation are compared with those from experiment.
The quantitative agreements between them indicate the following two facts: (i) the order of TNE considered in the current DBM is sufficient, (ii) the choosing of discrete velocities, spatial-temporal steps, and simulation parameters like the relaxation times are suitable for the following physical researches.
Then, five cases with various specific-heat ratios are simulated.
Several analysis methods for complex physical fields, including the description scheme of TNE behaviors, tracer particle method, and two-fluid  model, are used to characterize the effects of specific-heat ratio on the bubble shape, deformation process, average motion, vortex motion, mixing degree of the fluid system, TNE strength, and entropy production.
Specifically, for bubble shape, bubbles with different specific-heat ratios display various jet structures.
The smaller the specific-heat ratio is, the stouter the jet structure can be seen.
For the case with smaller specific-heat ratio, the fluid is easier to be compressed.
So, the characteristic scales of bubbles with smaller specific-heat ratio tend to be compressed smaller.
For the bubble, the smaller the specific-heat ratio, the slower average motion.
In the shock compression stage, the specific-heat ratio contributes little effects to the vortex motion. Differently, after the shock passes through the bubble, it significantly influences the vorticity around the interface and the corresponding amplitude of circulation due to the development of KHI.
The larger the difference in specific-heat ratio between the bubble and ambient gas, the higher the degree of material mixing.
Effects of specific-heat ratio on various TNE quantities are different.
These differences consistently show the complexity of TNE flows which is still far from a clear understanding.

In addition, it is found that the temporal evolution of the entropy production rates $\dot{S}_{\rm{NOMF}}$ and $\dot{S}_{\rm{NOEF}}$ both show three stages because of the influence of the shock wave location.
The smaller the specific-heat ratio,  the larger the velocity gradient, which indirectly enhances the strength of $\dot{S}_{\rm{NOMF}}$.
The specific-heat ratio increases the $\dot{S}_{\rm{NOEF}}$ by raising the temperature gradient.
The influence of specific-heat ratio on $S_{\rm{NOEF}}$ is more significant than that on $S_{\rm{NOMF}}$.
Effects of specific-heat ratio on entropy production caused by NOMF and NOEF are contrary.
Specifically, the entropy production contributed by NOMF increases with reduced specific-heat ratio.
But the entropy production caused by NOEF first reduces with decreasing specific-heat ratio and then approaches to a saturation value.
When the specific-heat ratio $\gamma$  is smaller than a threshold value $\gamma_c$ ($\gamma_c \approx 1.315$), the entropy production induced by NOEF is more significant than that caused by NOMF.
However, in the case of $\gamma>\gamma_c$, the situation reverses.
The fundamental research in this paper helps to understand the interaction mechanism between shock waves and bubbles in ICF, supersonic combustors, underwater explosions, etc.
The effects of viscosity and heat conduction on the interaction between shock waves and bubbles will be studied in the following work.

\section*{Acknowledgments}
The authors thank Chuandong Lin, Feng Chen, Ge Zhang, Yiming Shan, Jie Chen and Hanwei Li on helpful discussions on DBM.
The authors also thank Juchun Ding for providing experimental results.
This work was supported by the National Natural Science Foundation of China (under Grant Nos.  12172061, 11875001, and 12102397), the Strategic Priority Research Program of Chinese Academy of Sciences (under Grant No. XDA25051000), the opening project of State Key Laboratory of Explosion Science and Technology (Beijing Institute of Technology) (under Grant No. KFJJ23-16M), Foundation of Laboratory of Computational Physics, and Hebei Natural Science Foundation (Grant No. A2021409001), Central Guidance on Local Science and Technology Development Fund of Hebei Province (Grant No. 226Z7601G), and ``Three, Three and Three Talent Project'' of Hebei Province (Grant No. A202105005).

\appendix
\label{sec:Appendixes}
\section{ \textcolor{red}{Dimensionless variables versus actual physical variables }} \label{sec:AppendixesA}

All the physical variables used in simulation are dimensionless, so the actual physical quantities should be converted into dimensionless parameters.
The actual macroscopic quantities of SBI experiment in Ref. \cite{Ding2018POF} are as follows:
\[
\left\{
\begin{array}{l}
(\rho,u_x,u_y,p,T)^{\rm{bubble}}=( 4.859\rm{kg/m^3},0,0,101.3\rm{kPa},293\rm{K}) , \\
(\rho,u_x,u_y,p,T)^{\rm{Air}}=(1.2\rm{kg/m^3},0,0,101.3\rm{kPa},293\rm{K}) .
\end{array}
\right.
\]
Other physical quantities are: $R_{\rm{bubble}}=71.9\rm{J/(kg\cdot K)}$, $R_{\rm{Air}}=286.7\rm{J/(kg\cdot K)}$, $\gamma_{\rm{bubble}}=1.117$, and $\gamma_{\rm{Air}}=1.4$, $\mu_{\rm{Air}}=1.824\times10^{-5}\rm{Pa\cdot s}$.

The transformation relationship from physical parameters to dimensionless ones are as follows:
\begin{equation}
\hat{\rho}=\frac{\rho}{\rho_{\infty}},\hat{T}=\frac{T}{T_{\infty}},\hat{x}=\frac{x}{L_{\infty}},\hat{R}=\frac{R}{R_{\infty}},
\end{equation}
\begin{equation}
\hat{u}=\frac{u}{u_{\infty}},\hat{t}=\frac{t}{L_{\infty}/\sqrt{R_{\infty}T_{\infty}}},\hat{P}=\frac{P}{\rho_{\infty} R_{\infty} T_{\infty}},\hat{\mu}=\frac{\mu}{\rho_{\infty}L_{\infty}\sqrt{R_{\infty}T_{\infty}}},
\end{equation}
where $u_{\infty}=\sqrt{ R_{\infty} T_{\infty}}$.
In these equations, subscript ``$\infty$'' indicates the reference quantity and superscript ``$\wedge$'' means the dimensionless parameter.
In this part, we choose the macroscopic quantities in downstream region as the reference quantity, i.e., $\rho_{\infty}=\rho_{\rm{Air}}$, $T_{\infty}=T_{\rm{Air}}$, $L_{\infty}=2.917\rm{m}$, and $R_{\infty}=R_{\rm{Air}}$.

In this way, the dimensionless macroscopic quantities of the flow field are obtained:
\[
\left\{ \begin{gathered}
  (\rho,T,u_x ,u_y )_{\rm{0,bubble}}  = (4.0347,1.0,0.0,0.0), \hfill \\
  (\rho,T,u_x ,u_y )_{0,\rm{Air}}  = (1.0,1.0,0.0,0.0), \hfill \\
  (\rho,T,u_x ,u_y )_1  = (1.3416,1.128,0.3616,0.0), \hfill \\
\end{gathered}  \right.
\]
where the subscript ``0'' (``1'') represents downstream (upstream) region.
In the above equation, the initial parameters in upstream region (i.e., $(\rho,T,u_x ,u_y )_1$) are given by Rankine-Hugoniot conditions \cite{Zhang2020POF}.

\section{ \textcolor{red}{Expressions of the kinetic moments }} \label{sec:AppendixesB}
The expressions of kinetic moment used in our model are as follows:
\begin{equation}
M^{\sigma,eq}_{0}=\sum_{i}f^{\sigma,eq}_{i}=\rho^{\sigma} ,
\label{M0}
\end{equation}
\begin{equation}
M^{\sigma,eq}_{1,x}=\sum_{i}f^{\sigma,eq}_{i}v_{ix}=\rho^{\sigma}u_{x} ,
\end{equation}
\begin{equation}
M^{\sigma,eq}_{1,y}=\sum_{i}f^{\sigma,eq}_{i}v_{iy}=\rho^{\sigma}u_{y} ,
\end{equation}
\begin{eqnarray}
\begin{aligned}
M^{\sigma,eq}_{2,0}=\sum_{i}f^{\sigma,eq}_{i}(v_{i\alpha}^{2}+\eta^{\sigma2}_{i})
=\rho^{\sigma}[(D+I^{\sigma})R^{\sigma}T+u_{\alpha}^{2}]
 ,
\end{aligned}
\end{eqnarray}
\begin{equation}
M^{\sigma,eq}_{2,xy}=\sum_{i}f^{\sigma,eq}_{i}v_{ix}v_{iy}=\rho^{\sigma}u_{x}u_{y} ,
\end{equation}
\begin{equation}
M^{\sigma,eq}_{2,xx}=\sum_{i}f^{\sigma,eq}_{i}v_{ix}^{2}=\rho^{\sigma}u_{x}^{2} ,
\end{equation}
\begin{equation}
M^{\sigma,eq}_{2,yy}=\sum_{i}f^{\sigma,eq}_{i}v_{iy}^{2}=\rho^{\sigma}u_{y}^{2} ,
\end{equation}
\begin{eqnarray}
\begin{aligned}
M^{\sigma,eq}_{3,1,x}&=\sum_{i}f^{\sigma,eq}_{i}v_{ix}(v_{i\alpha}^{2}+\eta^{\sigma2}_{i})\\
&=\rho^{\sigma}u_{x}[(D+I^{\sigma}+2)R^{\sigma}T+u_{\alpha}^{2}]
 ,
\end{aligned}
\end{eqnarray}
\begin{eqnarray}
\begin{aligned}
M^{\sigma,eq}_{3,1,y}&=\sum_{i}f^{\sigma,eq}_{i}v_{iy}(v_{i\alpha}^{2}+\eta^{\sigma2}_{i})\\
&=\rho^{\sigma}u_{y}[(D+I^{\sigma}+2)R^{\sigma}T+u_{\alpha}^{2}]
 ,
\end{aligned}
\end{eqnarray}
\begin{equation}
M^{\sigma,eq}_{3,xxx}=\sum_{i}f^{\sigma,eq}_{i}v_{ix}^{3}=\rho^{\sigma}u_{x}(3R^{\sigma}T+u_{x}^{2}) ,
\end{equation}
\begin{equation}
M^{\sigma,eq}_{3,xxy}=\sum_{i}f^{\sigma,eq}_{i}v_{ix}^{2}v_{iy}=\rho^{\sigma}u_{y}(R^{\sigma}T+u_{x}^{2}) ,
\end{equation}
\begin{equation}
M^{\sigma,eq}_{3,xyy}=\sum_{i}f^{\sigma,eq}_{i}v_{ix}v_{iy}^{2}=\rho^{\sigma}u_{x}(R^{\sigma}T+u_{y}^{2}) ,
\end{equation}
\begin{equation}
M^{\sigma,eq}_{3,yyy}=\sum_{i}f^{\sigma,eq}_{i}v_{iy}^{3}=\rho^{\sigma}u_{y}(3R^{\sigma}T+u_{y}^{2}) ,
\end{equation}
\begin{eqnarray}
\begin{aligned}
M^{\sigma,eq}_{4,2,xx}&=\sum_{i}f^{\sigma,eq}_{i}v_{ix}^2(v_{i\alpha}^{2}+\eta^{\sigma2}_{i})=\rho^{\sigma} \{ (D+I^{\sigma}+2)R^{\sigma 2}T^2 \\
&+u_x^2(u_x^2+u_y^2) +R^{\sigma}T [  u_x^2(D+I^{\sigma}+5)+u_y^2 ] \}
,
\end{aligned}
\end{eqnarray}
\begin{eqnarray}
\begin{aligned}
M^{\sigma,eq}_{4,2,xy}&=\sum_{i}f^{\sigma,eq}_{i}v_{ix}v_{iy}(v_{i\alpha}^{2}+\eta^{\sigma2}_{i})\\
&=\rho^{\sigma} u_x u_y[(D+I^{\sigma}+4)R^{\sigma}T+u_x^2+u_y^2  ]
,
\end{aligned}
\end{eqnarray}
\begin{eqnarray}
\begin{aligned}
M^{\sigma,eq}_{4,2,yy}&=\sum_{i}f^{\sigma,eq}_{i}v_{iy}^2(v_{i\alpha}^{2}+\eta^{\sigma2}_{i})=\rho^{\sigma} \{ (D+I^{\sigma}+2)R^{\sigma 2}T^2 \\
&+u_y^2(u_x^2+u_y^2)+R^{\sigma}T [  u_y^2(D+I^{\sigma}+5)+u_x^2 ] \}
,
\label{M42yy}
\end{aligned}
\end{eqnarray}
where the subscript ``$m,n$'' means that the $m$-order tensor is contracted to $n$-order tensor.

\section{\textcolor{red}{Two-fluid hydrodynamic equations}} \label{sec:AppendixesC}
According to the CE multiscale analysis, the Boltzmann-BGK equation can be reduced to the hydrodynamic equations.
In the following part, the derivation process from Boltzmann-BGK equation to a two-fluid hydrodynamic equation are shown.
More details can see the reference presented by Zhang \emph{et al.} \cite{Zhang2020POF}.
The discrete Boltzmann equation for component $\sigma$ is
\begin{equation}
\frac{\partial f^{\sigma}_{i}}{\partial t}+v_{i\alpha}\cdot\frac{\partial f^{\sigma}_{i}}{\partial r_{\alpha}}=-\frac{1}{\tau^{\sigma}}(f^{\sigma}_{i}-f^{\sigma,seq}_{i})-\frac{1}{\tau^{\sigma}}(f^{\sigma,seq}_{i}-f^{\sigma,eq}_{i}),
\label{discrete-Boltzmann}
\end{equation}
In Eq. \ref{discrete-Boltzmann} there are two equilibrium distribution functions, i.e., $f^{\sigma,seq}=f^{\sigma,seq}(\rho^{\sigma},\mathbf{u}^{\sigma},T^{\sigma})$ and $f^{\sigma,eq}=f^{\sigma,eq}(\rho^{\sigma},\mathbf{u},T)$.
For convenience, $S^{\sigma}_{i}$ is defined as $S^{\sigma}_{i}=\frac{1}{\tau^{\sigma}}(f^{\sigma,seq}_{i}-f^{\sigma,eq}_{i})$.
We perform the CE expansion around the $f^{\sigma,seq}$.
That is, the distribution function $f^{\sigma}_{i}$ can be expanded as
\begin{equation}
f^{\sigma}_{i}=f^{\sigma,seq}_{i}+\epsilon f^{\sigma,(1)}_{i}+\epsilon^{2}f^{\sigma,(2)}_{i}+\cdots  ,
\end{equation}
where $\epsilon$ is a coefficient referring to Knudsen number.
The partial derivatives of time and space can also be expanded to
\begin{equation}
\frac{\partial}{\partial t}=\epsilon\frac{\partial}{\partial t_{1}}+\epsilon^{2}\frac{\partial}{\partial t}_{2}+\cdots  ,
\end{equation}
\begin{equation}
\frac{\partial}{\partial r_{\alpha}}=\epsilon\frac{\partial}{\partial r_{1\alpha}}  ,
\end{equation}
\begin{equation}
S_{i}^{\sigma}=\epsilon S_{1i}^{\sigma} ,
\end{equation}
Substituting the above four equations into Eq. (\ref{discrete-Boltzmann}), we can obtain
\begin{eqnarray}
\begin{aligned}
& \epsilon \frac{\partial (f^{\sigma,seq}_{i}+\epsilon f^{\sigma,(1)}_{i}+\epsilon^{2}f^{\sigma,(2)}_{i}+\cdots)}{\partial t_{1}} \\
& + \epsilon^2 \frac{\partial (f^{\sigma,seq}_{i}+\epsilon f^{\sigma,(1)}_{i}+\epsilon^{2}f^{\sigma,(2)}_{i}+\cdots)}{\partial t_{2}}\\
&+ \epsilon v_{i\alpha}\cdot\frac{\partial (f^{\sigma,seq}_{i}+\epsilon f^{\sigma,(1)}_{i}+\epsilon^{2}f^{\sigma,(2)}_{i}+\cdots)}{\partial r_{\alpha}}\\
&=-\frac{1}{\tau^{\sigma}}[(f^{\sigma,seq}_{i}+\epsilon f^{\sigma,(1)}_{i}+\epsilon^{2}f^{\sigma,(2)}_{i}+\cdots)-f^{\sigma,seq}_{i}]-\epsilon S_{1i}^{\sigma}.
\end{aligned}
\end{eqnarray}
When retaining to $\epsilon$ terms, the following equation is obtained
\begin{equation}
\epsilon \frac{\partial f^{\sigma,eq}_{i}}{\partial t_1}+\epsilon v_{i\alpha}\cdot\frac{\partial f^{\sigma,eq}_{i}}{\partial r_{1\alpha}}=-\epsilon\frac{1}{\tau^{\sigma}}f^{\sigma,(1)}_{i}-\epsilon S_{1i}^{\sigma},
\label{epsilon1}
\end{equation}
Performing operators $\sum_i$, $\sum_i v_{i\alpha}$, $\sum_i \frac{1}{2}(v_{i}^2+\eta_{i}^{2})$, we obtain
\begin{equation}
\epsilon \frac{\partial\rho^{\sigma}}{\partial t_1}+\epsilon \frac{\partial}{\partial r_{1\alpha}}(\rho^{\sigma} u_{\alpha}^{\sigma})=0
,
\label{Euler1}
\end{equation}
\begin{eqnarray}
\begin{aligned}
\epsilon \frac{\partial}{\partial t_1}(\rho^{\sigma} u_{\alpha}^{\sigma})
+\epsilon \frac{\partial (p^{\sigma}\delta_{\alpha\beta}+\rho^{\sigma}
u^{\sigma}_{\alpha}u^{\sigma}_{\beta})}{\partial r_{1\beta}}=-\epsilon\frac{\rho^{\sigma}}{\tau^{\sigma}}(u^{\sigma}_{\alpha}-u_{\alpha}),
\end{aligned}
\end{eqnarray}
\begin{eqnarray}
\begin{aligned}
\epsilon \frac{\partial}{\partial t_1}(\rho^{\sigma} E_{T}^{\sigma})&+\epsilon \frac{\partial}{\partial r_{1\alpha}}(\rho^{\sigma}E^{\sigma}_{T}+p^{\sigma})u^{\sigma}_{\alpha}\\
&=-\epsilon \frac{\rho^{\sigma}}{\tau^{\sigma}}[\frac{R^{\sigma}(D+I^{\sigma})(T^{\sigma}-T)}{2}+\frac{u_{\alpha}^{\sigma 2}-u_{\alpha}^{2}}{2}] ,
\label{Euler3}
\end{aligned}
\end{eqnarray}
where $p^{\sigma}=\rho^{\sigma}R^{\sigma}T^{\sigma}$ and $E^{\sigma}_{T}=\frac{1}{2}[(D+I^{\sigma})T+u^{\sigma 2}_{\alpha}]$.
In derivation, we need to use the kinetic moments of $\bm{M}_{m}^{\sigma,seq}$ and $\bm{M}_{m,n}^{\sigma,seq}$.
Their expressions can be obtained by replacing $u_{\alpha}$ and $T$ in Eqs. (\ref{M0}) to (\ref{M42yy}) of the Appendix B with $u_{\alpha}^{\sigma}$  and $T^{\sigma}$.
From the above three equations, we can obtain
\begin{equation}
\frac{\partial\rho^{\sigma}}{\partial t_1}= -\rho^{\sigma}\frac{\partial u_{\alpha}^{\sigma}}{\partial r_{1\alpha}}-u_{\alpha}^{\sigma}\frac{\partial \rho^{\sigma}}{\partial r_{1\alpha}}
,
\end{equation}
\begin{eqnarray}
\begin{aligned}
\frac{\partial u_{\alpha}^{\sigma} }{\partial t_1}
=-\frac{\partial T^{\sigma}}{\partial r_{1\alpha}}-\frac{T^{\sigma}}{\rho^{\sigma}}\frac{\partial \rho^{\sigma}}{\partial r_{1\alpha}}-u_{\beta}^{\alpha}\frac{\partial u_{\alpha}^{\sigma}}{\partial r_{1\beta}}+G
,
\end{aligned}
\end{eqnarray}
\begin{eqnarray}
\begin{aligned}
\frac{\partial T^{\sigma}}{\partial t_1}=-u^{\sigma}_{\alpha}\frac{\partial T^{\sigma}}{\partial r_{1\alpha}}-\frac{2T^{\sigma}}{D+I^{\sigma}}\frac{\partial u_{\alpha}^{\sigma}}{\partial r_{1\alpha}}+Z
,
\end{aligned}
\end{eqnarray}
where $G=-\frac{\rho^{\sigma}}{\tau^{\sigma}}(u^{\sigma}_{\alpha}-u_{\alpha})$ and $Z=-\frac{1}{\tau^{\sigma}}[(T^{\sigma}-T)-\frac{(u_{\alpha}^{\sigma}-u_{\alpha})^2}{D+I^{\sigma}}]$.

From Eq. (\ref{epsilon1}), we can obtain $f^{\sigma,(1)}_{i}$
\begin{equation}
f^{\sigma,(1)}_{i}=-\tau^{\sigma}(\frac{\partial f^{\sigma,eq}_{i}}{\partial t_1}+ v_{i\alpha}\cdot\frac{\partial f^{\sigma,eq}_{i}}{\partial r_{1\alpha}})-\tau^{\sigma} S_{1i}^{\sigma}.
\label{f1}
\end{equation}

When retaining to $\epsilon^2$ terms, we can obtain
\begin{equation}
\epsilon^2 \frac{\partial f^{\sigma,(1)}_{i}}{\partial t_1}+\epsilon^2 \frac{\partial f^{\sigma,seq}_{i}}{\partial t_2}+\epsilon^2 v_{i\alpha}\cdot\frac{\partial f^{\sigma,(1)}_{i}}{\partial r_{1\alpha}}=-\epsilon^2 \frac{1}{\tau^{\sigma}}f_{i}^{\sigma,(2)}
\label{f2}
\end{equation}
Substituting operators $\sum_i$, $\sum_i v_{i\alpha}$, $\sum_i \frac{1}{2}(v_{i}^2+\eta_{i}^{2})$ to Eq. (\ref{f2}), respectively, gives
\begin{equation}
\epsilon^2 \frac{\partial\rho^{\sigma}}{\partial t_2}= 0
,
\end{equation}
\begin{equation}
\epsilon^2 \frac{\partial (\rho^{\sigma} u_{\alpha}^{\sigma})}{\partial t_2}+\epsilon^2 \frac{\partial M_{2,\alpha\beta}(f^{\sigma,(1)})}{\partial r_{1\beta}} = 0
,
\end{equation}
\begin{equation}
\epsilon^2 \frac{\partial (\rho^{\sigma} E_{T}^{\sigma})}{\partial t_2}+\epsilon^2 \frac{\partial M_{3,1,\alpha}(f^{\sigma,(1)})}{\partial r_{1\alpha}} = 0
,
\end{equation}
where $ M_{2,\alpha\beta}(f^{\sigma,(1)}) $ $=$ $\sum_{i}v_{i\alpha}v_{i\beta}f_{i}^{\sigma,(1)}$ and $ M_{3,1,\alpha}(f^{\sigma,(1)}) $ $=$ $ \sum_{i}v_{i\alpha}(v_{i}^2+\eta_{i}^{2})f_{i}^{\sigma,(1)} $.
Substituting Eq. (\ref{f1}) into the above three equations, and replacing the time derivatives with the space derivatives, we obtain
\begin{equation}
\epsilon^2 \frac{\partial\rho^{\sigma}}{\partial t_2}= 0
,
\label{epsilon2-NS-1}
\end{equation}
\begin{equation}
\epsilon^2 \frac{\partial (\rho^{\sigma} u_{\alpha}^{\sigma})}{\partial t_2}+\epsilon^2 \frac{\partial (P^{\sigma}_{\alpha\beta}+U_{\alpha \beta}^{\sigma})}{\partial r_{\beta}} = 0
,
\end{equation}
\begin{equation}
\epsilon^2 \frac{\partial (\rho^{\sigma} E_{T}^{\sigma})}{\partial t_2}+\epsilon^2 \frac{\partial [u^{\sigma}_{\beta}(P^{\sigma}_{\alpha\beta}+U_{\alpha\beta}^{\sigma})-\kappa^{\sigma}\frac{\partial T^{\sigma}}{\partial r_{\alpha}}+Y_{\alpha}^{\sigma}]}{\partial r_{1\alpha}} = 0
,
\label{epsilon2-NS-3}
\end{equation}
where
\begin{equation}
P^{\sigma}_{\alpha\beta}=-\mu^{\sigma}(\frac{\partial u^{\sigma}_{\alpha}}{\partial r_{\beta}}+\frac{\partial u^{\sigma}_{\beta}}{\partial r_{\alpha}}-\frac{2}{D+I^{\sigma}}\frac{\partial u^{\sigma}_{\gamma}}{\partial r_{\gamma}}\delta_{\alpha\beta})
,
\end{equation}
\begin{equation}
U_{\alpha\beta}^{\sigma}=\rho^{\sigma}[(u_{\beta}-u^{\sigma}_{\beta})(u_{\alpha}-u^{\sigma}_{\alpha})
+\frac{1}{D+I^{\sigma}}(u^{\sigma}_{\alpha}-u_{\alpha})^{2}\delta_{\alpha\beta}]
,
\end{equation}
\begin{eqnarray}
\begin{aligned}
Y_{\alpha}^{\sigma}&=[\frac{D+I^{\sigma}+2}{2}\rho^{\sigma}R^{\sigma}(T^{\sigma}-T)(u^{\sigma}_{\alpha}-u_{\alpha})\\
&-\frac{D+I^{\sigma}+4}{2(D+I^{\sigma})}\rho^{\sigma}(u_{\alpha}^{\sigma}-u_{\alpha})^2u_{\alpha}^{\sigma}\\
&+\rho^{\sigma}u_{\alpha}^{\sigma}(u_{\alpha}^{\sigma}-u_{\alpha})u_{\alpha}^{\sigma}-\rho^{\sigma}u_{\beta}^{\sigma}(u_{\alpha}u_{\beta}-u_{\alpha}u_{\beta}^{\sigma})]\\
&+\frac{1}{2}\rho^{\sigma}(u^{2}_{\alpha}-u_{\alpha}^{\sigma 2})u_{\alpha}
.
\end{aligned}
\end{eqnarray}
Adding the equations for $\partial / \partial t_{1}$ (Eqs. (\ref{Euler1}) to (\ref{Euler3})) and $\partial / \partial t_{2}$ (Eqs. (\ref{epsilon2-NS-1}) to (\ref{epsilon2-NS-3})), the NS equations for component $\sigma$ are obtained,
\begin{equation}
\frac{\partial\rho^{\sigma}}{\partial t}+\frac{\partial}{\partial r_{\alpha}}(\rho^{\sigma} u_{\alpha}^{\sigma})=0
,
\end{equation}
\begin{eqnarray}
\begin{aligned}
\frac{\partial}{\partial t}(\rho^{\sigma} u_{\alpha}^{\sigma})
&+\frac{\partial (p^{\sigma}\delta_{\alpha\beta}+\rho^{\sigma}
u^{\sigma}_{\alpha}u^{\sigma}_{\beta})}{\partial r_{\beta}}\\
&+\frac{\partial (P^{\sigma}_{\alpha\beta}+U_{\alpha \beta}^{\sigma})}{\partial r_{\beta}}=0, \label{Eq:DDBM-N1}
\end{aligned}
\end{eqnarray}
\begin{eqnarray}
\begin{aligned}
&\frac{\partial}{\partial t}(\rho^{\sigma} E_{T}^{\sigma})+\frac{\partial}{\partial r_{\alpha}}(\rho^{\sigma}E^{\sigma}_{T}+p^{\sigma})u^{\sigma}_{\alpha}\\
&+\frac{\partial}{\partial r_{\beta}}[u^{\sigma}_{\beta}(P^{\sigma}_{\alpha\beta}+U_{\alpha\beta}^{\sigma})-\kappa^{\sigma}\frac{\partial T^{\sigma}}{\partial r_{\alpha}}+Y_{\alpha}^{\sigma}]=0 .
\end{aligned}
\end{eqnarray}

Performing $\sum_{\sigma}$ on both sides of the three equations gives the NS hydrodynamic equations describing the whole system,
\begin{equation}
\frac{\partial\rho}{\partial t}+\frac{\partial}{\partial r_{\alpha}}(\rho u_{\alpha})=0
,
\end{equation}
\begin{eqnarray}
\begin{aligned}
\frac{\partial}{\partial t}(\rho u_{\alpha})
&+\frac{\partial \sum\limits_{\sigma}(p^{\sigma}\delta_{\alpha\beta}+\rho^{\sigma}
u^{\sigma}_{\alpha}u^{\sigma}_{\beta})}{\partial r_{\beta}}\\
&+\frac{\partial\sum\limits_{\sigma} (P^{\sigma}_{\alpha\beta}+U_{\alpha \beta}^{\sigma})}{\partial r_{\beta}}=0, \label{Eq:DDBM-N1}
\end{aligned}
\end{eqnarray}
\begin{eqnarray}
\begin{aligned}
&\frac{\partial}{\partial t}(\rho E_{T})+\frac{\partial}{\partial r_{\alpha}}\sum\limits_{\sigma}(\rho^{\sigma}E^{\sigma}_{T}+p^{\sigma})u^{\sigma}_{\alpha}\\
&+\frac{\partial}{\partial r_{\beta}}\sum\limits_{\sigma}[u^{\sigma}_{\beta}(P^{\sigma}_{\alpha\beta}+U_{\alpha\beta}^{\sigma})-\kappa^{\sigma}\frac{\partial T^{\sigma}}{\partial r_{\alpha}}+Y_{\alpha}^{\sigma}]=0 .
\end{aligned}
\end{eqnarray}

The dynamic viscosity efficient is $\mu^{\sigma}=\tau^{\sigma} p^{\sigma}$, with $p^{\sigma}=\rho^{\sigma} R^{\sigma} T^{\sigma}$.
The heat conductivity is $\kappa^{\sigma}=C_{p}^{\sigma}\tau^{\sigma} p^{\sigma}$, with $C_{p}^{\sigma}=\frac{D+I^{\sigma}+2}{2}R^{\sigma}$.
For mixture, the expressions are
\begin{equation}
\mu=\tau p, \kappa=C_{p}\tau p
.
\end{equation}

It should be noted that the ability to recover the corresponding level of macroscopic fluid mechanics equations is only part of the physical function of DBM. The corresponding to the physical functions of DBM is the EHEs, which, in addition to the conserved moments evolution equations corresponding to the three conservation laws of mass, momentum and energy, also includes some of the most closely related non-conserved moments evolution equations.
We refer the EHEs derivation based on kinetic equation to as KMM.
 The necessity of the  expanded part, the evolution equations of the relevant non-conserved moments, increases rapidly as increasing the degree of non-continuity/non-equilibrium.
As the degree of non-continuity/non-equilibrium increases, the complexity will rapidly make KMM simulation studies, deriving and solving EHE,  impossible.



  \bibliographystyle{elsarticle-num}
  \bibliography{2-fluid-pof}





\end{document}